\documentclass[11pt,draftclsnofoot,onecolumn]{IEEEtran}

\usepackage{graphicx,slashbox}
\usepackage{epsfig}
\usepackage{amsmath}
\usepackage{bbm}
\usepackage{amssymb}
\usepackage{wrapfig}
\usepackage{times,color}
\usepackage{cite,footnote,xspace,syntonly,algorithm,algorithmic,bm}
\usepackage[caption=false,font=footnotesize]{subfig}

\usepackage{amsfonts}

\input{mysymbol.sty}
\newcommand{\calN}{{\cal N}}

\newcommand{\calS}{{\cal S}}
\newcommand{\calJ}{{\cal J}}

\def \cN {\mathcal{N}}
\def \cL {\mathcal{L}}

\def \cS {\mathcal{S}}

\def \cP {\mathcal{P}}
\def \cJ {\mathcal{J}}

\def \bT {{\bf T}}
\def \bY {{\bf Y}}
\def \bR {{\bf R}}

\def \bX {{\bf X}}
\def \bA {{\bf A}}
\def \bV {{\bf V}}
\def \bW {{\bf W}}
\def \bF {{\bf F}}
\def \bG {{\bf G}}
\def \bI {{\bf I}}
\def \bB {{\bf B}}
\def \bM {{\bf M}}
\def \bU {{\bf U}}
\def \bV {{\bf V}}
\def \bE {{\bf E}}
\def \bD {{\bf D}}
\def \bZ {{\bf Z}}
\def \bW {{\bf W}}

\def \bL {{\bf L}}
\def \bQ {{\bf Q}}
\def \bC {{\bf C}}
\def \bO {{\bf O}}
\def \bP {{\bf P}}

\def \bv {{\bf v}}
\def \bx {{\bf x}}

\def \bv {{\bf v}}

\def \bSigma {{\bf \Sigma}}

\def \tr {\text{tr}}
\def \vec {\text{vec}}
\def \supp {\text{supp}}

\def \vec {\text{vec}}
\def \unvec {\text{unvec}}

\long\def\symbolfootnote[#1]#2{\begingroup
\def\thefootnote{\fnsymbol{footnote}}
\footnote[#1]{#2}\endgroup} \psfull

\begin{document}
\title{\huge In-network Sparsity-regularized Rank Minimization: Algorithms and Applications$^\dag$}

\author{{\it Morteza Mardani, Gonzalo~Mateos, and Georgios~B.~Giannakis~(contact author)$^\ast$}}

\markboth{IEEE TRANSACTIONS ON SIGNAL PROCESSING (SUBMITTED)}
\maketitle \maketitle \symbolfootnote[0]{$\dag$ Work in this paper
was supported by the MURI Grant No. AFOSR FA9550-10-1-0567. Parts of the
paper appeared in the {\it Proc. of the 45th Asilomar Conference on Signals, Systems,
and Computers}, Pacific Grove, CA, Nov. 6-9, 2011, and were submitted to the
{\it 13th Workshop on Signal Processing Advances in Wireless Communications},
Cesme, Turkey, Jun. 17-20, 2012.} \symbolfootnote[0]{$\ast$ The authors are with the Dept.
of Electrical and Computer Engineering, University of
Minnesota, 200 Union Street SE, Minneapolis, MN 55455. Tel/fax:
(612)626-7781/625-4583; Emails:
\texttt{\{morteza,mate0058,georgios\}@umn.edu}}

\vspace*{-80pt}
\begin{center}
\small{\bf Submitted: }\today\\
\end{center}


\thispagestyle{empty}\addtocounter{page}{-1}
\begin{abstract}
Given a limited number of entries from the superposition of a low-rank matrix plus the product of a
known fat compression matrix times a sparse matrix, recovery of
the low-rank and sparse components is a fundamental task subsuming compressed
sensing, matrix completion, and principal components pursuit. This paper develops algorithms for \textit{distributed} sparsity-regularized
rank minimization over networks, when the nuclear- and $\ell_1$-norm are used
as surrogates to the rank and nonzero entry counts of the sought
matrices, respectively. While nuclear-norm minimization has well-documented merits when
centralized processing is viable, non-separability of the singular-value sum challenges its distributed minimization.
To overcome this limitation, an alternative characterization
of the nuclear norm is adopted which leads to a
separable, yet non-convex cost minimized via the alternating-direction
method of multipliers. The novel distributed iterations entail
reduced-complexity per-node tasks, and affordable message passing among single-hop neighbors. Interestingly, upon convergence the
distributed (non-convex) estimator provably attains the global
optimum of its centralized counterpart, regardless of initialization.
Several application domains are outlined to highlight
the generality and impact of the  proposed framework. These include
unveiling traffic anomalies in backbone networks, predicting networkwide path
latencies, and mapping the RF ambiance using wireless
cognitive radios. Simulations with synthetic and real network data corroborate the
convergence of the novel distributed algorithm, and its centralized performance guarantees.
\end{abstract}
\vspace*{-5pt}
\begin{center} \bfseries EDICS Category: SEN-DIST, SPC-APPL, SEN-COLB. \end{center}
%
\newpage


\section{Introduction}\label{sec:intro}

Let $\bX:=[x_{l,t}]\in\mathbb{R}^{L\times T}$ be a \textit{low-rank} matrix
[$\textrm{rank}(\bX)\ll \min(L,T)$], and $\bA:=[a_{f,t}]\in\mathbb{R}^{F\times T}$
be a \textit{sparse} matrix with support size considerably smaller than $F T$.
Consider also a matrix $\bR:=[r_{l,t}]\in\mathbb{R}^{L\times F}$ and a set
$\Omega\subseteq \{1,\ldots,L\}\times \{1,\ldots,T\}$ of index pairs
$(l,t)$ that define a sampling of the entries of $\bX$.
Given $\bR$ and a number of (possibly) noise corrupted measurements
\begin{align}
y_{l,t} = x_{l,t} + \sum_{f=1}^Fr_{l,f} a_{f,t} + v_{l,t}, \quad (l,t)\in\Omega   \label{eq:y_lt}
\end{align}
the goal is to estimate low-rank $\bX$ and sparse $\bA$,
by denoising the observed entries and imputing the missing ones.
Introducing the sampling operator $\cP_{\Omega}(\cdot)$ which sets the
entries of its matrix argument not in $\Omega$ to zero and leaves the rest unchanged,
the data model can be compactly written in matrix form as
\begin{align}
\cP_{\Omega}(\bY) = \cP_{\Omega}(\bX + \bR\bA + \bV). \label{eq:model}
\end{align}
A natural estimator accounting for the low rank of $\bX$ and the sparsity of $\bA$ will be sought to fit the data
$\cP_{\Omega}(\bY)$ in the least-squares (LS) error sense, as well as minimize the
rank of $\bX$, and the number of nonzero entries of $\bA$ measured by its $\ell_0$-(pseudo)
norm; see e.g.~\cite{CP10},~\cite{MMG12},~\cite{CLMW09},~\cite{CSPW11}
for related problems subsumed by the one described
here. Unfortunately, both rank and $\ell_0$-norm minimization
are in general NP-hard problems~\cite{Natarajan_NP_duro,rank_NP_Duro}.
Typically, the nuclear norm $\|\bX\|_*:=\sum_{k}\sigma_{k}(\bX)$
($\sigma_{k}(\bX)$ denotes the $k$-th singular value of $\bX$)
and the $\ell_1$-norm $\|\bA\|_1:=\sum_{f,t}|a_{f,t}|$
are adopted as surrogates,
since they are the closest \textit{convex}
approximants to $\textrm{rank}(\bX)$ and $\|\bA\|_0$, respectively~\cite{RFP07,CT05,tropp06tit}.
Accordingly, one solves
\begin{align}
\text{(P1)}~~~~~\min_{\{\bX,\bA\}} ~& \frac{1}{2}\|\cP_{\Omega}
(\bY - \bX - \bR\bA)\|_F^2 + \lambda_* \|\bX\|_{*} + \lambda_1 \|\bA\|_1 \nonumber
\end{align}
where $\lambda_*,\lambda_1\geq 0$ are rank- and sparsity-controlling parameters.
Being convex (P1) is appealing, and some of its special instances
are known to attain good performance in theory and practice.
For instance,
when no data are missing (P1) can be used to unveil traffic anomalies in networks~\cite{MMG12}.
Results in~\cite{MMG12} show that $\bX$ and $\bA$ can be exactly recovered
in the absence of noise,
even when $\bR$ is a fat (compression) operator. When $\bR$ equals the identity matrix,
(P1) reduces to the so-termed robust principal component analysis (PCA),
for which exact recovery results are available in~\cite{CLMW09} and~\cite{CSPW11}. Moreover, for the special case $\bR \equiv \mathbf{0}_{L \times F}$, (P1) offers a low-rank matrix completion alternative with well-documented merits; see
e.g.,~\cite{CR08} and~\cite{CP10}. Stable recovery results in the presence of noise
are also available for matrix completion and robust PCA~\cite{CP10,zlwcm10}. Earlier
efforts dealing with the recovery of sparse vectors in noise led to
similar performance guarantees; see e.g.,~\cite{bickel09}.

In all these works, the samples $\cP_{\Omega}(\bY)$ and matrix $\bR$ are assumed centrally available, so that they can be jointly processed to estimate $\bX$ and $\bA$ by e.g., solving (P1). Collecting all this information can be
challenging though in various applications of interest, or may be even impossible
in e.g., wireless sensor networks (WSNs) operating under stringent power budget
constraints. In other cases such as the Internet or collaborative marketing studies,
agents providing private data for e.g., fitting a low-rank preference
model, may not be willing to share their training data but only the learning results.
Performing the optimization in a centralized fashion raises robustness concerns as
well, since the central processor represents an isolated point of failure.
Scalability is yet another driving force motivating distributed solutions.
Several customized iterative algorithms have been
proposed to solve instances of (P1), and have been shown effective
in tackling low- to medium-size problems; see e.g.,~\cite{MMG12},~\cite{CR08},~\cite{RFP07}.
However, most algorithms require computation of singular values
per iteration and become prohibitively expensive when dealing with
high-dimensional data~\cite{Recht_Parallel_2011}. All in all, the aforementioned
reasons motivate the reduced-complexity \textit{distributed} algorithm for nuclear and
$\ell_1$-norm minimization developed in this paper.

In a similar vein, stochastic gradient algorithms were recently
developed for large-scale problems entailing regularization with the nuclear norm~\cite{Recht_Parallel_2011}.
Even though iterations in~\cite{Recht_Parallel_2011} are highly paralellizable,
they are not applicable to networks of arbitrary topology. There are also several
studies on distributed estimation of sparse signals via $\ell_1$-norm regularized
regression; see e.g.,~\cite{mateos_dlasso,dusan11,chouvardas11}.
Different from the treatment here,
the data model of~\cite{mateos_dlasso} is devoid of a low-rank component
and all the observations $\bY$ are assumed available (but distributed
across several interconnected agents). Formally, the model therein
is a special case of \eqref{eq:model} with $T=1$, $\bX=\mathbf{0}_{L\times T}$, and
$\Omega = \{1,\ldots,L\}\times \{1,\ldots,T\}$, in which case (P1) boils down
to finding the least-absolute shrinkage and selection operator (Lasso)~\cite{tibshirani_lasso}.

Building on the general model \eqref{eq:model} and the centralized estimator (P1),
this paper develops decentralized algorithms to estimate low-rank and sparse
matrices, based on in-network processing of a small subset of noise-corrupted
 and spatially-distributed measurements (Section \ref{sec:dist_sol}).
This is a challenging task however, since the
non-separable nuclear-norm present in (P1) is not amenable to distributed
minimization. To overcome this limitation, an alternative characterization of
the nuclear norm is adopted in Section \ref{ssec:separable},
which leads to a separable yet non-convex cost that
is minimized via the alternating-direction method of multipliers (AD-MoM)~\cite{Bertsekas_Book_Distr}.
The novel distributed iterations entail reduced-complexity optimization
subtasks per agent, and affordable
message passing only between single-hop neighbors (Section \ref{ssec:ad_mom}).
Interestingly, the
distributed (non-convex) estimator upon convergence provably attains {\it the global} optimum of its
centralized counterpart (P1), regardless of initialization. To demonstrate
the generality of the proposed estimator and its algorithmic framework, four
networking-related application domains are outlined  in Section \ref{sec:appl},
namely: i) unveiling traffic volume anomalies for large-scale networks~\cite{LCD04,MMG12};
ii) robust PCA~\cite{CSPW11,CLMW09}; iii) low-rank matrix
completion for networkwide path latency prediction~\cite{latencyprediction},
and iv) spectrum sensing for cognitive
radio (CR) networks~\cite{mateos_dlasso,juan_sensig}. Numerical tests with synthetic and real
network data drawn from these application domains corroborate the effectiveness and convergence
of the novel distributed algorithms, as well as their centralized performance benchmarks (Section \ref{sec:sims}).

Section \ref{sec:disc} concludes the paper,
while several technical details are deferred to the Appendix.

\noindent{\it Notation}: Bold uppercase (lowercase) letters will denote
matrices (column vectors), and calligraphic letters will be used for sets.
Operators $(\cdot)'$, $\rm{tr}(\cdot)$, $\sigma_{\max}(\cdot)$, $\odot$ and $\otimes$,
will denote transposition,
matrix trace, maximum singular value, Hadamard product, and Kronecker product, respectively;
$|\cdot|$ will be used for the cardinality of a set, and the magnitude of a scalar.
The matrix
vectorization operator $\vec(\bZ)$ stacks the columns of matrix $\bZ$
on top of each other to return a supervector, and its inverse is
$\unvec(\mathbf{z})$.
The diagonal matrix $\rm{diag(\bv)}$ has the entries of $\bv$ on its diagonal,
and the positive semidefinite matrix $\mathbf{M}$ will be
denoted by $\bbM\succeq\mathbf{0}$.
The $\ell_p$-norm of $\bx \in \mathbb{R}^n$ is $\|\bx\|_p:=(\sum_{i=1}^n |x_i|^p)^{1/p}$
for $p \geq 1$. For two matrices $\bM,\bU \in \mathbb{R}^{n \times n}$,
$\langle \bM, \bU \rangle := \rm{tr(\bM' \bU)}$ denotes their trace inner product.
The Frobenious norm of matrix $\bM = [m_{i,j}] \in \mathbb{R}^{n \times p}$ is
$\|\bM\|_F:=\sqrt{\tr(\bM\bM')}$,
$\|\bM\|:=\max_{\|\bx\|_2=1} \|\bM\bx\|_2$ is the spectral norm,
$\|\bM\|_1:=\sum_{i,j} |m_{i,j}|$ is the $\ell_1$-norm,
$\|\bM\|_{\infty}:=\max_{i,j} |m_{i,j}|$ is the $\ell_{\infty}$-norm,
and $\|\bM\|_{\ast}:=\sum_{i}\sigma_i(\bM)$ is the nuclear norm, where
$\sigma_i(\bM)$ denotes the $i$-th singular value of $\bM$. The $n \times n$ identity
matrix will be represented by $\bI_n$, while $\mathbf{0}_{n}$ will stand for $n \times 1$ vector of all zeros, and $\mathbf{0}_{n \times p}:=\mathbf{0}_{n}
\mathbf{0}'_{p}$. Similar notations will be adopted for vectors (matrices)
of all ones.


\section{Preliminaries and Problem Statement}
\label{sec:model}
Consider $N$ networked agents capable of performing some
local computations, as well as exchanging messages among
directly connected neighbors. An agent should be understood as
an abstract entity, e.g., a sensor in a WSN, a router
monitoring Internet traffic; or a
sensing CR from a next-generation communications
technology. The network is
modeled as an undirected graph $G(\cN,\cL)$, where the set of nodes
$\cN:=\{1,\ldots,N\}$ corresponds to the network agents, and the edges
(links) in $\cL:=\{1,\ldots,L\}$ represent pairs of agents that can communicate.
Agent $n\in\cN$ communicates with its single-hop neighboring peers
in $\cJ_n$, and the size of the neighborhood will be henceforth
denoted by $|\cJ_n|$. To
ensure that the data from an arbitrary agent can eventually
percolate through the entire network, it is assumed that:
\begin{description}
\item [{\bf (a1)}] \emph{Graph $G$ is connected; i.e., there
exists a (possibly) multi-hop path connecting any
two agents.}
\end{description}

With reference to the low-rank and sparse matrix recovery problem outlined in
Section~\ref{sec:intro}, in the network setting envisioned here each
agent $n\in\cN$ acquires a few incomplete and noise-corrupted
rows of matrix $\bY$. Specifically, the local data available to
agent $n$ is matrix $\cP_{\Omega_n}(\bY_n)$, where $\bY_n\in\mathbb{R}^{L_n\times T}$,
$\sum_{n=1}^N L_n=L$, and $\bY:=\left[\bbY_1^\prime,\ldots,\bbY_N^\prime\right]^\prime=
\bX+\bR\bA+\bV$.
The index pairs in $\Omega_n$ are those in $\Omega$
for which the row index matches the rows of $\bY$ observed by agent $n$.
Additionally, suppose that agent $n$ has available the local matrix
$\bR_n \in\mathbbm{R}^{L_n \times F}$, containing a row subset of $\bR$
associated with the observed rows in $\bY_n$, i.e, $\bR:=\left[\bbR_1^\prime,\ldots
,\bbR_N^\prime\right]^\prime$. Agents collaborate to form
the wanted estimator (P1) in a distributed fashion, which can be
equivalently rewritten as
\begin{align}
\min_{\{\bX,\bA\}}&~\sum_{n=1}^N\left[\frac{1}{2}\|\cP_{\Omega_n}(\bY_n - \bX_n - \bR_n\bA)\|_F^2 +
\frac{\lambda_*}{N}\|\bX\|_{\ast} + \frac{\lambda_1}{N} \|\bA\|_1 \right]\nonumber.
\end{align}
The objective of this paper is to develop a distributed algorithm for
sparsity-regularized rank minimization via (P1), based on in-network processing of the locally
available data. The described setup naturally suggests three characteristics
that the algorithm should exhibit: c1) agent $n\in\cN$ should obtain an
estimate of $\bbX_n$ and $\bA$, which coincides with the corresponding
solution of the centralized estimator (P1) that uses the entire data
$\cP_{\Omega}(\bY)$; c2) processing per agent should be kept as simple
as possible; and c3) the overhead for inter-agent communications should be affordable
and confined to single-hop neighborhoods.


\section{Distributed Algorithm for In-Network Operation}
\label{sec:dist_sol}
To facilitate reducing the computational complexity and memory storage
requirements of the distributed algorithm sought, it is henceforth assumed
that:
\begin{description}
\item [{\bf (a2)}] \emph{ An upper bound $\rho \geq\textrm{rank}(\hat\bbX)$ on the rank of matrix $\hat\bX$ obtained via (P1) is available. }
\end{description}
As argued next, the smaller the value of $\rho$, the more efficient the
algorithm becomes. A small value of $\rho$ is well motivated in various
applications. For example, the Internet traffic analysis of backbone
networks in~\cite{LCD04} demonstrates that origin-to-destination flows
have a very low intrinsic dimensionality, which renders the traffic
matrix low rank; see also Section \ref{ssec:unveil_anomaly}.
In addition, recall that $\textrm{rank}(\hat\bbX)$ is controlled
by the choice of $\lambda_\ast$ in (P1), and the
rank of the solution can be made small enough, for
sufficiently large $\lambda_\ast$. Because $\textrm{rank}(\hat\bbX)\leq \rho$,
(P1)'s search space is effectively reduced and one can factorize the
decision variable as $\bX=\bL\bQ'$, where $\bL$ and $\bQ$ are
$L \times \rho$ and $T \times \rho$ matrices, respectively.
Adopting this reparametrization of $\bX$ in (P1), one obtains
the following equivalent optimization problem
\begin{align}
\text{(P2)}~~~~~\min_{\{\bL,\bQ,\bA\}}& \sum_{n=1}^N\left[\frac{1}{2}\|\cP_{\Omega_n}(\bY_n
- \bL_n\bQ' - \bR_n\bA)\|_{F}^{2} +
\frac{\lambda_{\ast}}{N}\|\bL\bQ'\|_{\ast}+\frac{\lambda_1}{N}\|\bA\|_1\right]\nonumber
\end{align}
which is non-convex due to the bilinear terms $\bL_n\bQ'$, and
$\bbL:=\left[\bbL_1^\prime,\ldots,\bbL_N^\prime\right]^\prime$.
The number of variables is reduced from $LT+FT$ in (P1),
to $\rho(L+T)+FT$ in (P2). The savings can be significant when $\rho$
is in the order of a few dozens, and both $L$ and $T$ are large.
The dominant $FT$-term in the variable count of (P3) is due
to $\bbA$, which is sparse and can be efficiently handled even when
both $F$ and $T$ are large.
Problem (P3) is still not amenable to distributed implementation due to: (i)
the non-separable nuclear norm present in the cost function; and (ii) the global
variables $\bbQ$ and $\bbA$ coupling the per-agent summands.


\subsection{A separable nuclear norm regularization}\label{ssec:separable}
To address (i), consider the following neat characterization
of the nuclear norm~\cite{RFP07,Recht_Parallel_2011}
\begin{equation}\label{eq:nuc_nrom_def}
\|\bX\|_*:=\min_{\{\bL,\bQ\}}~~~ \frac{1}{2}\left\{\|\bL\|_F^2+\|\bQ\|_F^2 \right\},\quad
\text{s. to}~~~ \bX=\bL\bQ'.
\end{equation}
For an arbitrary matrix $\bX$ with SVD $\bX=\bU_X\bSigma_X\bV_X'$,
the minimum in \eqref{eq:nuc_nrom_def}
is attained for $\bL=\bU_X\bSigma_X^{1/2}$ and $\bQ=\bV_X\bSigma_X^{1/2}$.
The optimization \eqref{eq:nuc_nrom_def} is over all possible bilinear factorizations
of $\bbX$, so that the number of columns of $\mathbf{L}$ and $\mathbf{Q}$
is also a variable. Leveraging~\eqref{eq:nuc_nrom_def}, the following reformulation
of (P2) provides an important first step towards obtaining a distributed estimator:
%
\begin{align}
\text{(P3)}~~~~~\min_{\{\bL,\bQ,\bA\}}& \sum_{n=1}^N\left[\frac{1}{2}\|\cP_{\Omega_n}(\bY_n
- \bL_n\bQ' - \bR_n\bA)\|_{F}^{2} +
\frac{\lambda_{*}}{2N}\left\{N\|\bL_n\|_F^2 + \|\bQ\|_F^2 \right\}+
\frac{\lambda_1}{N}\|\bA\|_1\right]. \nonumber
\end{align}
As asserted in the following lemma, adopting the
separable Frobenius-norm regularization in (P3) comes
with no loss of optimality, provided the upper bound $\rho$ is chosen large enough.

\begin{lemma}\label{lem:lem_1}
Under (a2), (P3) is equivalent to (P1).
\end{lemma}

\begin{IEEEproof}
Let $\{\hat\bX,\hat\bA\}$ denote the minimizer of (P1). Clearly, $\textrm{rank}(\hat\bX) \leq \rho$,
implies that (P2) is equivalent to (P1). From \eqref{eq:nuc_nrom_def} one can also
infer that for every feasible solution $\{\bL,\bQ,\bA\}$, the cost in (P3) is no
smaller than that of (P2). The gap between the globally minimum costs of (P2)
and (P3) vanishes at $\bar{\bA}:=\hat\bA$, $\bar{\bL}:=\hat{\bU} \hat{\bSigma}^{1/2}$,
and $\bar{\bQ}:=\hat{\bV}\hat{\bSigma}^{1/2}$, where $\hat\bX=\hat{\bU}\hat{\bSigma}\hat{\bV}'$.
Therefore, the cost functions of (P1) and (P3) are identical at the minimum.
\end{IEEEproof}

Lemma~\ref{lem:lem_1} ensures that by finding the global minimum of (P3)
[which could have significantly less variables than (P1)], one can  recover the optimal solution of
(P1). However, since (P3) is non-convex, it may have stationary points which need not
be globally optimum. Interestingly, the next proposition shows that under
relatively mild assumptions on $\textrm{rank}(\hat\bX)$
and the noise variance, every stationary point of (P3) is globally optimum for (P1).
For a proof, see Appendix A.

\begin{proposition}\label{prop:prop_1}
Let $\{\bar{\bL},\bar{\bQ},\bar{\bA}\}$ be a stationary point of (P3).
If $\|\cP_{\Omega}(\bY-\bar{\bL}\bar{\bQ}'-\bR\bar{\bA})\| \leq \lambda_*$~(no subscript in $\|.\|$ signifies spectral norm),
then $\{\hat\bX=\bar{\bL}\bar{\bQ}',\hat\bA=\bar{\bA}\}$ is the globally optimal solution of (P1).
\end{proposition}

\noindent Condition $\|\cP_{\Omega}(\bY-\bar{\bL}\bar{\bQ}'-\bR\bar{\bA})\| \leq \lambda_*$
captures tacitly the role of $\rho$, the number of columns of $\bL$
and $\bQ$ in the postulated model. In particular, for sufficiently small $\rho$ the residual
$\|\cP_{\Omega}(\bY-\bar{\bL}\bar{\bQ}'-\bR\bar{\bA})\| $
becomes large and consequently the condition is violated (unless $\lambda_*$ is large
enough, in which case a sufficiently low-rank solution to (P1) is expected). This is manifested through the fact that the extra condition
$\textrm{rank}(\hat\bX) \leq \rho$ in Lemma \ref{lem:lem_1} is no longer needed.
In addition, note that the noise variance certainly affects the
value of $\|\cP_{\Omega}(\bY-\bar{\bL}\bar{\bQ}'-\bR\bar{\bA})\|$,
and thus satisfaction of the said condition.


\subsection{Local variables and consensus constraints}
\label{ssec:localvar}
To decompose the cost function in (P3), in which summands are coupled through
the global variables $\bbQ$ and $\bbA$ [cf. (ii) at the beginning of this
section], introduce auxiliary variables $\{\bbQ_n,\bbA_n\}_{n=1}^N$
representing local estimates of $\{\bbQ,\bbA\}$ per agent $n$. These local
estimates are utilized to form the separable \emph{constrained} minimization
problem
\begin{align}
\text{(P4)}~~~~~\min_{\{\bL,\bQ_n,\bA_n,\bB_n\}}&
\sum_{n=1}^N\left[\frac{1}{2}\|\cP_{\Omega_n}(\bY_n
- \bL_n\bQ_n' - \bR_n\bB_n)\|_{F}^{2} +
\frac{\lambda_{*}}{2N}\left\{N\|\bL_n\|_F^2 + \|\bQ_n\|_F^2 \right\}+
\frac{\lambda_1}{N}\|\bA_n\|_1\right] \nonumber\\
\text{s. to} &\quad \bB_n=\bA_n,\quad n\in\mathcal{N}\nonumber\\
&\quad\bQ_n=\bQ_m,\:\quad\bA_n=\bA_m, \quad
m\in\cJ_n,\:n\in\mathcal{N}.\nonumber
\end{align}
For reasons that will become clear later on, additional variables
$\{\bB_n\}_{n=1}^N$ were introduced to split the $\ell_2-$norm fitting-error
part of
the cost of (P4), from the $\ell_1-$norm regularization on the $\{\bA_n\}_{n=1}^N$
(cf. Remark \ref{remark:general_sparse}).
These extra variables are not needed if $\bR'\bR=\bI_F$.
The set of additional constraints $\bB_n=\bA_n$ ensures that, in this sense, nothing
changes in going from (P3) to (P4).
Most importantly, (P3) and (P4) are equivalent
optimization problems under (a1).
The equivalence should be understood in the sense that
$\hat{\bQ}_1=\hat{\bQ}_2=\ldots=\hat{\bQ}_N=\hat{\bQ}$ and likewise for $\bA$,
where $\{\hat{\bQ}_n,\hat{\bA}_n\}_{n\in\cN}$ and $\{\hat{\bQ},\hat{\bA}\}$ are the
optimal solutions of (P4) and (P3), respectively. Of course, the corresponding
estimates of $\bL$ will coincide as well. Even though consensus is a
fortiori imposed within neighborhoods, it extends to the whole (connected)
network and local estimates agree on the global solution of (P3). To arrive
at the desired distributed algorithm, it is convenient to reparametrize
the consensus constraints in (P4) as
\begin{align}
\bQ_n={}&{}\bar{\bF}_n^m,\:\bQ_m=\tilde{\bF}_n^m, \textrm{ and }
\bar{\bF}_n^m=\tilde{\bF}_n^m,  \quad
m\in\cJ_n,\:n\in\mathcal{N}\label{eq:constr_1}\\
\bA_n={}&{}\bar{\bG}_n^m,\:\bA_m=\tilde{\bG}_n^m, \textrm{ and }
\bar{\bG}_n^m=\tilde{\bG}_n^m,  \quad
m\in\cJ_n,\:n\in\mathcal{N}\label{eq:constr_2}
\end{align}
where
$\{\bar{\bF}_n^m,\tilde{\bF}_n^m,\bar{\bG}_n^m,\tilde{\bG}_n^m\}^{m\in\cJ_n}_{n\in\cN}$ are auxiliary optimization variables that will be eventually eliminated.


\subsection{The alternating-direction method of multipliers}
\label{ssec:ad_mom}
To tackle the constrained minimization problem (P4), associate
Lagrange multipliers $\bM_n$ with the splitting
constraints $\bB_n=\bA_n$, $n\in\cN$. Likewise,
associate additional dual variables $\bar{\bC}_n^m$ and
$\tilde{\bC}_n^m$ ($\bar{\bD}_n^m$ and $\tilde{\bD}_n^m$)
with the first  pair of consensus constraints
in \eqref{eq:constr_1} [respectively \eqref{eq:constr_2}]. Next introduce the quadratically \textit{augmented} Lagrangian  function
\begin{align}
\ccalL_c\left(\mathcal{V}_1,\mathcal{V}_2,\mathcal{V}_3,\mathcal{M}\right){}={}&
\sum_{n=1}^N\left[\frac{1}{2}\|\cP_{\Omega_n}(\bY_n
- \bL_n\bQ_n' - \bR_n\bB_n)\|_{F}^{2} +
\frac{\lambda_{*}}{2N}\left\{N\|\bL_n\|_F^2 + \|\bQ_n\|_F^2 \right\}+
\frac{\lambda_1}{N}\|\bA_n\|_1\right]\nonumber
\end{align}
\begin{align}\label{augLagr}
& \hspace{1cm}+\sum_{n=1}^N\langle \bM_n,\bB_n-\bA_n\rangle
+\frac{c}{2}\sum_{n=1}^N\|\bB_n-\bA_n\|_F^2\nonumber\\
& \hspace{1cm}+\sum_{n=1}^N\sum_{m\in\calJ_n}\left\{\langle \bar{\bC}_n^m,\bQ_n-\bar{\bF}_n^m\rangle
+\langle \tilde{\bC}_n^m,\bQ_m-\tilde{\bF}_n^m\rangle+
\langle \bar{\bD}_n^m,\bA_n-\bar{\bG}_n^m\rangle+\langle \tilde{\bD}_n^m,\bA_m-\tilde{\bG}_n^m\rangle
\right\}\nonumber\\
&\hspace{1cm}+\frac{c}{2}\sum_{n=1}^N\sum_{m\in\calJ_n}\left\{\|\bQ_n-\bar{\bF}_n^m\|_F^2
+\|\bQ_m-\tilde{\bF}_n^m\|_F^2+
\|\bA_n-\bar{\bG}_n^m\|_F^2+\|\bA_m-\tilde{\bG}_n^m\|_F^2
\right\}
\end{align}
where $c$ is a positive penalty coefficient, and the primal variables
are split into three groups
$\mathcal{V}_1:=\{\bQ_n,\bA_n\}_{n=1}^{N}$, $\mathcal{V}_2:=\{\bL_n\}_{n=1}^{N}$,
and $\mathcal{V}_3:=\{\bB_n,\bar{\bF}_n^m,\tilde{\bF}_n^m,\bar{\bG}_n^m,\tilde{\bG}_n^m\}_{n\in\cN}^{m\in\cJ_n}$.
For notational convenience, collect all multipliers in
$\mathcal{M}:=\{\bM_n,\bar{\bC}_n^m,\tilde{\bC}_n^m,\bar{\bD}_n^m,\tilde{\bD}_n^m\}_{n\in\cN}^{m\in\cJ_n}$.
Note that the remaining constraints in \eqref{eq:constr_1} and \eqref{eq:constr_2},
namely $C_V:=\{\bar{\bF}_n^m=\tilde{\bF}_n^m,\;\bar{\bG}_n^m=\tilde{\bG}_n^m, \:m\in\cJ_n,\:n\in\mathcal{N}\}$,
have not been dualized.

To minimize (P4) in a distributed fashion, a variation of the
alternating-direction method of multipliers (AD-MoM) will be
adopted here. The AD-MoM is an iterative augmented Lagrangian
method especially well-suited for parallel processing~\cite{Bertsekas_Book_Distr},
which has been proven successful to tackle the optimization tasks
encountered e.g., with distributed estimation problems~\cite{Yannis_Ale_GG_PartI,mateos_dlasso}.
The proposed solver entails an iterative procedure comprising four
steps per iteration $k=1,2,\ldots$
\begin{description}
\item [{\bf [S1]}] \textbf{Update dual variables:}\begin{align}
    \bM_n[k]&=\bM_n[k-1]+\mu(\bB_n[k]-\bA_n[k]),{\quad}n\in\cN\label{eq:multi_M}\\
    \bar{\bC}_n^m[k]&=\bar{\bC}_n^m[k-1]+\mu(\bQ_n[k]-\bar{\bF}_n^m[k]),
    {\quad}n\in\cN,\:m\in\cJ_n\label{eq:multi_barC}\\
    \tilde{\bC}_n^m[k]&=\tilde{\bC}_n^m[k-1]+\mu(\bQ_m[k]-\tilde{\bF}_n^m[k]),
    {\quad}n\in\cN,\:m\in\cJ_n\label{eq:multi_tildeC}\\
    \bar{\bD}_n^m[k]&=\bar{\bD}_n^m[k-1]+\mu(\bA_n[k]-\bar{\bG}_n^m[k]),
    {\quad}n\in\cN,\:m\in\cJ_n\label{eq:multi_barD}\\
    \tilde{\bD}_n^m[k]&=\tilde{\bD}_n^m[k-1]+\mu(\bA_m[k]-\tilde{\bG}_n^m[k]),
    {\quad}n\in\cN,\:m\in\cJ_n\label{eq:multi_tildeD}
\end{align}

\item [{\bf [S2]}]  \textbf{Update first group of primal variables:}
    \begin{equation}\mathcal{V}_1[k+1]=
    \mbox{arg}\:\min_{\mathcal{V}_1}\ccalL_c\left(\mathcal{V}_1,\mathcal{V}_2[k],
    \mathcal{V}_3[k],\mathcal{M}[k]\right).
    \label{S2_ADMOM}\end{equation}

\item [{\bf [S3]}]  \textbf{Update second group of primal variables:}
        \begin{equation}\mathcal{V}_2[k+1]=
            \mbox{arg}\:\min_{\mathcal{V}_2}\ccalL_c\left(\mathcal{V}_1[k+1],\mathcal{V}_2,
            \mathcal{V}_3[k],\mathcal{M}[k]\right).
        \label{S3_ADMOM}\end{equation}

\item [{\bf [S4]}] \textbf{Update auxiliary primal variables:}
    \begin{equation}\mathcal{V}_3[k+1]=
        \mbox{arg}\:\min_{\mathcal{V}_3\in C_V}\ccalL_c\left(\mathcal{V}_1[k+1],\mathcal{V}_2[k+1],
        \mathcal{V}_3,\mathcal{M}[k]\right).
    \label{S4_ADMOM}\end{equation}
\end{description}
This four-step procedure implements a block-coordinate descent method with dual variable updates. At each step while minimizing the augmented
Lagrangian, the variables not being updated are treated as fixed
and are substituted with their most up to date values. Different
from AD-MoM, the alternating-minimization step
here generally cycles over three groups of primal variables
$\mathcal{V}_1$-$\mathcal{V}_3$ (cf. two groups in
AD-MoM~\cite{Bertsekas_Book_Distr}). In some special instances
detailed in Sections~\ref{ssec:matrixcompletion} and \ref{ssec:dlasso},
cycling over two groups of variables only is sufficient. In [S1], $\mu>0$ is the step size of the subgradient ascent
iterations
\eqref{eq:multi_M}-\eqref{eq:multi_tildeD}. While it is
common in AD-MoM implementations to select $\mu=c$, a distinction
between the step size and the penalty parameter is made explicit
here in the interest of generality.

Reformulating the estimator (P1) to its equivalent form (P4) renders
the augmented Lagrangian in \eqref{augLagr} highly decomposable.
The separability comes in two flavors, both with respect to the
variable groups $\mathcal{V}_1$, $\mathcal{V}_2$, and $\mathcal{V}_3$,
as well as across the network agents $n\in\cN$. This in turn leads to
highly parallelized, simplified recursions corresponding to the
aforementioned four steps. Specifically, it is shown in Appendix
B that if the multipliers are initialized to zero, [S1]-[S4]
constitute the distributed algorithm tabulated under Algorithm \ref{tab:table_1}.
For conciseness in presenting the algorithm, define the local residuals
$r_n(\bL_n,\bQ_n,\bB_n):=\frac{1}{2}\|\cP_{\Omega_n}(\bY_n - \bL_n\bQ_n' - \bR_n\bB_n)\|_{F}^{2}$.
In addition, define the soft-thresholding matrix
$\cS_{\tau}(\bM)$ with $(i,j)$-th entry given by
$\textrm{sign}(m_{i,j})\max\{|m_{i,j}|-\tau,0\}$, where $m_{i,j}$ denotes the $(i,j)$-th entry of $\bM$.

\begin{algorithm}[t]
\caption{: AD-MoM solver per agent $n\in\cN$} \small{
\begin{algorithmic}
	\STATE \textbf{input} $\bY_n, \bR_n, \lambda_{*}, \lambda_1, c, \mu$
    \STATE \textbf{initialize}
$\bM_n[0]=\bP_n[0]=\bA_n[1]=\bB_n[1]=\mathbf{0}_{F\times T}$,
$\bO[0]=\mathbf{0}_{T\times \rho}$, and $\bL_n[1],\:\bQ_n[1]$ at random
    \FOR {$k=1,2$,$\ldots$}
        \STATE Receive $\{\bQ_m[k],\bA_m[k]\}$ from neighbors $m\in\cJ_n$
        \STATE {\bf [S1]} \textbf{Update local dual variables:}
                \STATE $\bM_n[k]=\bM_n[k-1]+\mu (\bB_n[k]-\bA_n[k])$
                \STATE
$\bO_n[k]=\bO_n[k-1]+\mu \sum_{m\in\cJ_n}(\bQ_n[k]-\bQ_m[k])$
                \STATE
$\bP_n[k]=\bP_n[k-1]+\mu \sum_{m\in\cJ_n}(\bA_n[k]-\bA_m[k])$
        \STATE {\bf [S2]}  \textbf{Update first group of local primal variables:}
        \STATE
$\bQ_n[k+1]=\arg\min_{\bQ}\left\{
r_n(\bL_n[k],\bQ,\bB_n[k]) +\frac{\lambda_{*}}{2N}\|\bQ\|_F^2+
\langle \bO_n[k],\bQ \rangle +
c\sum_{m\in\cJ_n} \left\|\bQ-\frac{\bQ_n[k]+\bQ_m[k]}{2}\right\|_F^{2}\right\}$
        \STATE $\bA_n[k+1]= [c(1+2|\cJ_n|)]^{-1} \cS_{\lambda_1/N} \left(\bM_n[k]+c\bB_n[k]-
        \bP_n[k]+\hspace{-0.2mm}c\sum_{m\in\cJ_m}\hspace{-0.5mm}(\bA_n[k]+\bA_m[k])\right)$
        \STATE {\bf [S3]}  \textbf{Update second group of local primal variables:}
        \STATE
$\bL_n[k+1]=\arg\min_{\bL}\left\{r_n(\bL,\bQ_n[k+1],\bB_n[k]) +
\frac{\lambda_{*}}{2}\|\bL\|_F^2 \right\}$
		\STATE {\bf [S4]}  \textbf{Update auxiliary local primal variables:}
        \STATE $\bB_n[k+1]{}={}\arg\min_{\bB}\left\{r_n(\bL_n[k+1],\bQ_n[k+1],\bB)
        +\langle \bM_n[k],\bB\rangle+
        \frac{c}{2}\|\bB-\bA_n[k+1]\|_F^2\right\}$
        \STATE Broadcast $\{\bQ_n[k+1],\bA_n[k+1]\}$ to neighbors $m\in\cJ_n$
    \ENDFOR
    \RETURN $\bA_n, \bQ_n, \bL_n$
\end{algorithmic}}
\label{tab:table_1}
\end{algorithm}

\begin{remark}[Simplification of redundant variables]\label{remark:simplif} Careful
inspection of Algorithm \ref{tab:table_1} reveals that the inherently
redundant auxiliary variables and multipliers
$\{\bar{\bF}_n^m,\tilde{\bF}_n^m,\bar{\bG}_n^m,\tilde{\bG}_n^m,\tilde{\bC}_n^m,\tilde{\bD}_n^m\}$
have been eliminated. Agent $n$ does not need to
\textit{separately} keep track of all its non-redundant multipliers
$\{\bar{\bC}_n^m,\bar{\bD}_n^m\}_{m\in\cJ_n}$, but only to
update their respective (scaled) sums $\bO_n[k]:=2\sum_{m\in\cJ_n}\bar{\bC}_n^m[k]$ and
$\bP_n[k]:=2\sum_{m\in\cJ_n}\bar{\bD}_n^m[k]$.
\end{remark}

\begin{remark}[Computational and communication cost]\label{remark:cost}
The main computational burden of the algorithm stems from solving
unconstrained quadratic programs locally to update ${\bQ_n,\bL_n,\bB_n}$, and
to carry out simple soft-thresholding operations to update $\bA_n$.
On a per iteration basis, network agents communicate their
updated local estimates $\{\bQ_n[k],\bA_n[k]\}$ with their neighbors,
to carry out the updates of the primal and dual
variables during the next iteration. Regarding communication cost, $\bQ_n[k]$
is a $T \times \rho$ matrix and its transmission does not incur significant
overhead when $\rho$ is small. In
addition, the $F \times T$ matrix $\bA_n[k]$ is sparse, and can be
communicated efficiently. Observe that the dual variables need not be exchanged.
\end{remark}

\begin{remark}[General sparsity-promoting regularization]\label{remark:general_sparse}
Even though $\lambda_1\|\bA\|_1$ was adopted in (P1) to encourage sparsity
in the entries of $\bA$, the algorithmic framework here can accommodate more general
\textit{structured sparsity}-promoting penalties $\psi(\bA)$. To maintain
the per-agent computational complexity at affordable levels, the minimum requirement
on the admissible penalties is that the \textit{proximal operator}
\begin{equation}
\textrm{prox}_{\psi}(\tilde{\bY}):=\arg\min_{\bA}\left[\frac{1}{2}\|\tilde{\bY}-\bA\|_F^2+\psi(\bA)\right] \label{eq:proximal}
\end{equation}
is given in terms of vector or (and) scalar soft-thresholding operators.
In addition to the $\ell_1$-norm (Lasso penalty), this holds for the
sum of row-wise $\ell_2$-norms (group Lasso penalty~\cite{yuan_group_lasso}),
or, a linear combination of the aforementioned two -- the so-termed hierarchical
Lasso penalty that encourages sparsity across and within the rows of $\bA$~\cite{hilasso}.
All this is possible since by introducing the cost-splitting variables $\bB_n$,
the local sparse matrix updates are $\bA_n[k+1]=\textrm{prox}_{\psi}(\tilde{\bY}_n[k])$
for suitable $\tilde{\bY}_n[k]$ (see Appendix B).
\end{remark}

When employed to solve non-convex problems such as (P4), AD-MoM
(or its variant used here) offers no convergence guarantees. However, there is ample
experimental evidence in the literature which supports convergence
of AD-MoM, especially when the non-convex problem at hand exhibits
``favorable'' structure. For instance, (P4) is bi-convex and gives
rise to the strictly convex optimization subproblems
\eqref{S2_ADMOM}-\eqref{S4_ADMOM}, which admit unique closed-form
solutions per iteration. This observation and the linearity of the
constraints endow Algorithm~\ref{tab:table_1} with good convergence properties -- extensive numerical tests including those
presented in Section~\ref{sec:sims} demonstrate that this is
indeed the case. While a formal convergence proof goes beyond the scope of this paper, the following proposition proved
in Appendix C asserts that upon convergence, Algorithm
\ref{tab:table_1} attains consensus and global optimality.

\begin{proposition}\label{th:th_1}
If the sequence of iterates $\{\bQ_n[k],\bL_n[k],\bA_n[k]\}_{n\in\cN}$ generated by
Algorithm \ref{tab:table_1} converge to
$\{\bar{\bQ}_n,\bar{\bL}_n,\bar{\bA}_n\}_{n\in\cN}$, and (a1) holds, then: i)~$\bar{\bQ}_n=\bar{\bQ}_m,~\bar{\bA}_n=\bar{\bA}_m,~n,m\in\cN$;
and ii) if $\|\cP_{\Omega}(\bY-\bar{\bL}\bar{\bQ}_1'-\bR\bar{\bA}_1)\| \leq \lambda_*$,
then $\hat{\bX}=\bar{\bL}{\bar{\bQ}_1}'$ and $\hat{\bA}=\bar{\bA}_1$, where
$\{\hat{\bA},\hat{\bX}\}$ is the global optimum of (P1).
\end{proposition}

\section{Applications}
\label{sec:appl}
This section outlines a few applications that could benefit from the distributed sparsity-regularized rank minimization framework
described so far. In each case, the problem statement calls
for estimating low-rank $\bX$ and (or) sparse $\bA$,
given distributed data adhering to an application-dependent model subsumed by \eqref{eq:model}.
Customized algorithms are thus obtained as special cases
of the general iterations in Algorithm \ref{tab:table_1}.


\subsection{Unveiling traffic anomalies in backbone networks}
\label{ssec:unveil_anomaly}
In the backbone of large-scale networks, origin-to-destination (OD) traffic
flows experience abrupt changes which can result in congestion, and
limit the quality of service provisioning of the end users. These so-termed
\emph{traffic volume anomalies} could be due to external sources such as
network failures, denial of service attacks, or, intruders which hijack the
network services~\cite{MC03}~\cite{LCD04}. Unveiling such
anomalies is a crucial task towards engineering network traffic. This is a
challenging task however, since the available data are usually high-dimensional
noisy link-load measurements, which comprise the superposition of
\emph{unobservable} OD flows as explained next.

The network is modeled as in Section \ref{sec:model}, and
transports a set of end-to-end flows $\cal{F}$ (with $|\mathcal{F}| := F$) associated
with specific source-destinations pairs. For backbone
networks, the number of network layer
flows is typically much larger than the number of physical links $(F \gg L)$. Single-path
routing is considered here to send the traffic flow from a source to its intended
destination. Accordingly, for a particular
flow multiple links connecting the corresponding source-destination pair are
chosen to carry the traffic. Sparing details that can
be found in~\cite{MMG12}, the traffic $\bbY:=[y_{l,t}]\in\mathbb{R}^{L\times T}$
carried over links $l\in\cL$ and measured at time instants $t\in [1,T]$
can be compactly expressed as
\begin{equation}
\bY=\bR \left(\bZ + \bA\right) + \bV \label{eq:Y}
\end{equation}
where the fat routing matrix $\bbR:=[r_{\ell,f}]\in\{0,1\}^{L\times F}$
is fixed and given, $\bbZ:=[z_{f,t}]$ denotes the unknown ``clean''
traffic flows over the time horizon of
interest, and $\bbA:=[a_{f,t}]$ collects the traffic volume anomalies. These data
are distributed.
Agent $n$ acquires a few rows of $\bbY$ corresponding to the
link-load traffic measurements $\bbY_n
\in\mathbb{R}^{L_n\times T}$
from its outgoing links, and has available its local routing
table $\bR_n$ which indicates the OD flows routed through $n$.
Assuming a suitable ordering of links, the per agent quantities relate
to their global counterparts in \eqref{eq:Y} through
$\bY:=\left[\bbY_1^\prime,\ldots,\bbY_N^\prime\right]^\prime$ and
$\bR:=\left[\bbR_1^\prime,\ldots,\bbR_N^\prime\right]^\prime$.


%
%

Common temporal patterns among the traffic
flows in addition to their periodic behavior, render most rows (respectively columns)
of $\bZ$ linearly dependent, and thus $\bZ$
typically has  low rank~\cite{LCD04}. Anomalies are expected to occur sporadically
over time, and only last for short periods relative to the (possibly long)
measurement interval $[1,T]$. In addition, only a small fraction of the flows are
supposed to be anomalous at any given
time instant. This renders the anomaly matrix $\bA$ sparse across
rows and columns. Given local measurements $\{\bY_n\}_{n\in\cN}$
and the routing tables $\{\bR_n\}_{n\in\cN}$,
the goal is to estimate $\bA$ in a distributed fashion,
by capitalizing on the sparsity of $\bA$ and the low-rank property of $\bZ$. Since the primary goal is to recover $\bA$, define $\bX:=\bR\bZ$ which inherits
the low-rank property from $\bZ$, and consider [cf.~\eqref{eq:Y}]
\begin{align}
\bY= \bX + \bR \bA + \bV \label{eq:Y_modf}.
\end{align}
Model~\eqref{eq:Y_modf} is a special case of
\eqref{eq:model}, when all the entries of $\bY$ are observed, i.e.,
$\Omega = \{1,\ldots,L\}\times \{1,\ldots,T\}$. Note that $\bR\bA$ is not
sparse even though $\bA$ is itself sparse, hence principal
components pursuit is not applicable here~\cite{zlwcm10}.
Instead, the following estimator is adopted to unveil network anomalies~\cite{MMG12}
\begin{align}
\{\hat{\bX},\hat{\bA}\} = \arg\min_{\{\bX,\bA\}} \sum_{n=1}^N\left[ \frac{1}{2}
\|\bY_n-\bX_n-\bR_n\bA\|_F^2 + \frac{\lambda_{\ast}}{N} \|\bX\|_{\ast} + \frac{\lambda_{1}}{N} \|\bA\|_1 \right] \nonumber
\end{align}
which is subsumed by (P1). Accordingly, a distributed algorithm
can be readily obtained by simplifying the general iterations under Algorithm
\ref{tab:table_1}, the subject dealt with next.

\noindent\textbf{Distributed Algorithm for Unveiling Network Anomalies (DUNA).}
For the specific case here in which $\Omega = \{1,\ldots,L\}\times \{1,\ldots,T\}$,
the residuals in Algorithm~\ref{tab:table_1} reduce to
$r_n(\bL_n,\bQ_n,\bB_n):=\frac{1}{2}\|\bY_n - \bL_n\bQ_n' - \bR_n\bB_n\|_{F}^{2}$.
Accordingly, to update the primal variables $\bQ_n[k+1]$, $\bL_n[k+1]$
and $\bB_n[k+1]$ as per Algorithm~\ref{tab:table_1}, one needs to solve
respective unconstrained strictly convex quadratic optimization problems.
These admit closed-form solutions
detailed under Algorithm~\ref{tab:table_2}. The DUNA updates of the local
anomaly matrices $\bA_n[k+1]$ are given in terms of soft-thresholding operations,
exactly as in Algorithm \ref{tab:table_1}.

\begin{algorithm}[t]
\caption{: DUNA per agent $n\in\cN$} \small{ \label{tab:table_2}
\begin{algorithmic}
	\STATE \textbf{input} $\bY_n, \bR_n, \lambda_{*}, \lambda_1, c, \mu$
    \STATE \textbf{initialize}
$\bM_n[0]=\bP_n[0]=\bA_n[1]=\bB_n[1]=\mathbf{0}_{F\times T}$,
$\bO[0]=\mathbf{0}_{T\times \rho}$, and $\bL_n[1],\:\bQ_n[1]$ at random
    \FOR {$k=1,2$,$\ldots$}
        \STATE Receive $\{\bQ_m[k],\bA_m[k]\}$ from neighbors $m\in\cJ_n$
        \STATE {\bf [S1]} \textbf{Update local dual variables:}
                \STATE $\bM_n[k]=\bM_n[k-1]+\mu(\bB_n[k]-\bA_n[k])$
                \STATE $\bO_n[k]=\bO_n[k-1]+\mu\sum_{m\in\cJ_n}(\bQ_n[k]-\bQ_m[k])$
                \STATE $\bP_n[k]=\bP_n[k-1]+\mu\sum_{m\in\cJ_n}(\bA_n[k]-\bA_m[k])$
        \STATE {\bf [S2]}  \textbf{Update first group of local primal variables:}
        \STATE$\bQ_n[k+1]=\left\{\bY_n'\bL_n[k]- \bB_n'[k]\bR_n'\bL_n[k]-\bO_n[k]+ c\sum_{m\in\cJ_n} (\bQ_n[k]+\bQ_m[k])\right\} \left[\bL_n'[k]\bL_n[k]+(\lambda_*/N+2c|\cJ_n|)\bI_\rho\right]^{-1}$
        \STATE $\bA_n[k+1]=[c(1+2|\cJ_n|)]^{-1} \cS_{\lambda_1/N}
\left(\bM_n[k]+c\bB_n[k]-\bP_n[k]+\hspace{-0.2mm}c\sum_{m\in\cJ_m}\hspace{-0.5mm}(\bA_n[k]+\bA_m[k])\right)$
		\STATE {\bf [S3]}  \textbf{Update second group of local primal variables:}
        \STATE $\bL_n[k+1]=\left(\bY_n-\bR_n\bB_n[k]\right)\bQ_n[k+1]\left[\bQ^\prime_n[k+1]
\bQ_n[k+1]+\lambda_*\bI_\rho\right]^{-1}$
		\STATE {\bf [S4]}  \textbf{Update auxiliary local primal variables:}
        \STATE $\bB_n[k+1]=[\bR_n' \bR_n + c\bI_F]^{-1} \left\{\bR_n'(\bY_n - \bL_n[k+1] \bQ_n'[k+1]) - \bM_n[k]+
c\bA_n[k+1] \right\}$
        \STATE Broadcast $\{\bQ_n[k+1],\bA_n[k+1]\}$ to neighbors $m\in\cJ_n$
    \ENDFOR
    \RETURN $\bA_n, \bQ_n, \bL_n$
\end{algorithmic}}
\label{tab:table_2}
\end{algorithm}

Conceivably, the number of flows $F$ can be quite large,
thus inverting the $F\times F$ matrix
$\bR_n^\prime \bR_n + c\bI_F$ to update $\bB_n[k+1]$ could be complex
computationally. Fortunately, the inversion needs to be carried
out once, and can be performed and cached off-line. In addition, to reduce the inversion cost, the SVD of the local routing matrices
 $\bR_n=\bU_{R_n} \bSigma_{R_n} \bV_{R_n}^\prime$ can be obtained first,
and the matrix inversion lemma can be subsequently employed
to obtain $[\bR_n^\prime \bR_n + c\bI_F]^{-1}=(1/c)\left[\bI_p - \bV_{R_n} \bC
\bV_{R_n}^\prime\right]$, where $\bC:=\mathrm{diag}
\left(\frac{\sigma_1^2}{c+\sigma_1^2},...,\frac{\sigma_p^2}{c+\sigma_p^2}
\right)$ and $p= {\rm rank} (\bR_n)\ll F$. This computational shortcut
is commonly adopted in statistical learning algorithms when ridge regression
estimates are sought, and the number of variables is much larger
than the number of elements in the training set~\cite[Ch. 18]{elements_of_statistics}.
During the operational phase of the algorithm,
the main computational burden of DUNA comes from
repeated inversions of (small) $\rho\times\rho$ matrices, and parallel
soft-thresholding operations. The communication overhead is identical to
the one incurred by Algorithm \ref{tab:table_1} (cf. Remark \ref{remark:cost}).

\begin{remark}[Incomplete link traffic measurements]\label{remark:missing_meas}
In general, one can allow for missing
traffic data and the DUNA updates are still expressible in closed form.
\end{remark}


\subsection{In-network robust principal component analysis}
\label{subsec:RPCA}
Principal component analysis (PCA) is the workhorse of
high-dimensional data analysis and dimensionality reduction, with
numerous applications in statistics, networking, engineering, and
the biobehavioral sciences; see, e.g.,~\cite{PCA}.
Nowadays ubiquitous e-commerce sites, complex networks such as the Web, and urban traffic
surveillance systems generate massive volumes of data. As a result,
extracting the most informative, yet
low-dimensional structure from high-dimensional datasets is of
paramount importance~\cite{elements_of_statistics}.

Data obeying postulated low-rank models include also outliers, which are
samples not adhering to those nominal models. Unfortunately,
similar to LS estimates PCA is very sensitive to the outliers~\cite{PCA}.
While robust approaches to PCA are available, recently polynomial-time algorithms with remarkable performance guarantees
have emerged for low-rank matrix recovery in the presence of sparse --
but otherwise arbitrarily large -- errors~\cite{zlwcm10,CLMW09,CSPW11}.
Robust PCA is of great interest in networking-related applications.
One can think of distributed estimation using reduced-dimensionality sensor
observations~\cite{wsncompress}, and unveiling anomalous flows in backbone
networks from Netflow data~\cite{netflowpcp}; see also Section \ref{subsec:sims_rpca}.

In the network setting of Section \ref{sec:model}, each agent
$n \in\cN$ acquires $F_n$  outlier-plus-noise
corrupted rows of matrix $\bY:=[\bY_1',\ldots ,\bY_N']'$, where $\sum_{n=1}^N F_n=F$.
Local data can thus be modeled as $\bY_n=\bX_n+\bA_n+\bV_n$, where
$\bX:=[\bX_1',\ldots ,\bX_N']'$ has low rank. Agents want to estimate $\bX_n$
(and the outliers $\bA_n$) in a distributed fashion by forming the
global estimator~\cite{zlwcm10}
\begin{align}
\{\hat{\bX},\hat{\bA}\} = \arg\min_{\{\bX,\bA\}} \sum_{n=1}^N\left[ \frac{1}{2}
\|\bY_n-\bX_n-\bA_n\|_F^2 + \frac{\lambda_{\ast}}{N} \|\bX\|_{\ast} + \lambda_1 \|\bA_n\|_1 \right]
\label{eq:rpca}
\end{align}
which is once more a special case of (P1) when $\bR=\bI_F$. 

\noindent\textbf{Distributed Robust Principal Component Analysis (DRPCA) Algorithm.}
Regarding the general distributed formulation in (P4), the first constraint is no longer needed since $\bR=\bI_F$ [cf. the discussion after (P4)].
As agent $n$ is interested in estimating $\bA_n$ and $\|\bA\|_1$
is separable over the rows of $\bA$, the only required
constraints are $\bQ_n=\bQ_m,~m\in\cJ_n,~n\in\cN$. These are associated with the dual variables
$\bO_n$ per agent, and are updated according to Algorithm~\ref{tab:table_3}.
All in all, each agent stores and recursively updates the primal variables $\{\bQ_n,\bL_n\}$,
along with the $F_n \times T$ matrix $\bA_n$.

Mimicking the procedure that led to Algorithm \ref{tab:table_1},
one finds that primal variable updates in DRPCA are expressible in closed form.
In particular, the local outlier matrix $\bA_n[k+1]$ minimizes the Lasso cost
\begin{align}
\bA_n[k+1] = \arg\min_{\{\bA_n\}} \left\{ \frac{1}{2}
\|\bY_n-\bL_n[k+1]\bQ_n'[k+1]-\bA_n\|_F^2 + \lambda_1 \|\bA_n\|_1 \right\}
\nonumber
\end{align}
and is given in terms of soft-thresholding operations
as seen in Algorithm~\ref{tab:table_3}
[observe that $\bA_n[k+1] =\textrm{prox}_{\|\cdot\|_1}(\bY_n-\bL_n[k+1]\bQ_n'[k+1])$,
where $\textrm{prox}_{\psi}(\cdot)$ is defined in~\eqref{eq:proximal}].
\begin{algorithm}[t]
\caption{: DRPCA algorithm per agent $n\in\cN$} \small{ \label{tab:table_3}
\begin{algorithmic}
	\STATE \textbf{input} $\bY_n, \lambda_{*}, \lambda_1, c, \mu$
    \STATE \textbf{initialize}
$\bA_n[1]=\mathbf{0}_{F_n\times T}$,
$\bO[0]=\mathbf{0}_{T\times \rho}$, and $\bL_n[1],\:\bQ_n[1]$ at random.
    \FOR {$k=1,2$,$\ldots$}
        \STATE Receive $\{\bQ_m[k]\}$ from neighbors $m\in\cJ_n$
        \STATE {\bf [S1]} \textbf{Update local dual variables:}
                \STATE$\bO_n[k]=\bO_n[k-1]+\mu\sum_{m\in\cJ_n}(\bQ_n[k]-\bQ_m[k])$

        \STATE {\bf [S2]}  \textbf{Update first group of local primal variables:}
        \STATE$\bQ_n[k+1]=\left\{\bY_n'\bL_n[k]-
\bA_n'[k]\bL_n[k]-\bO_n[k]+ c\sum_{m\in\cJ_n} (\bQ_n[k]+\bQ_m[k])\right\} \left[\bL_n'[k]\bL_n[k]+(\lambda_*/N+2c|\cJ_n|)\bI_\rho\right]^{-1}$

		\STATE {\bf [S2]}  \textbf{Update second group of local primal variables:}
        \STATE
$\bL_n[k+1]=\left(\bY_n-\bA_n[k]\right)\bQ_n[k+1]\left[\bQ^\prime_n[k+1]
\bQ_n[k+1]+\lambda_*\bI_\rho\right]^{-1}$
\STATE {\bf [S3]}  \textbf{Update third group of local primal variables:}
\STATE $\bA_n[k+1]= \cS_{\lambda_1} \left( \bY_n-\bL_n[k+1] \bQ_n'[k+1]\right)$

        \STATE Broadcast $\{\bQ_n[k+1]\}$ to neighbors $m\in\cJ_n$

    \ENDFOR
    \RETURN $\bA_n, \bQ_n, \bL_n$
\end{algorithmic}}
\label{tab:table_3}
\end{algorithm}
DRPCA iterations are simple with small $\rho \times \rho$ matrices inverted
per iteration to update $\bL_n$ and
$\bQ_n$ (see Algorithm~\ref{tab:table_3}). Regarding communication
cost, each agent only broadcasts a $T \times \rho$ matrix $\bQ_n$
to its neighbors.


\subsection{Distributed low-rank matrix completion}
\label{ssec:matrixcompletion}
The ability to recover a low-rank matrix from a subset of
its entries is the leitmotif of recent advances for localization
of wireless sensors~\cite{mo10}, Internet traffic analysis~\cite{latencyprediction},~\cite{zrwq09},
and preference modeling for recommender
systems~\cite{gedas}.
In the \textit{low-rank matrix completion} problem, given
a limited number of (possibly) noise corrupted entries
of a low-rank matrix $\bX$, the goal is
to recover the entire matrix while denoising the observed
entries, and accurately imputing the missing ones.

In the network setting envisioned here, agent $n\in\cN$ has available $L_n$
incomplete and noise-corrupted rows of $\bY:=[\bY_1',\ldots ,\bY_N']'$.
Local data can thus be modeled as
$\cP_{\Omega_n}(\bY_n) = \cP_{\Omega_n}(\bX_n + \bV_n)$.
Relying on in-network processing,
agents aim at completing their own rows by forming the global estimator
\begin{align}
\hat{\bX}=\arg\min_{\bX}  \sum_{n=1}^N\left[
\frac{1}{2}\|\cP_{\Omega_n}(\bY_n - \bX_n)\|_F^2 +
\frac{\lambda_{\ast}}{N} \|\bX\|_{\ast} \right] \label{eq:mc_problem}
\end{align}
which exploits the low-rank property of $\bX$ through nuclear-norm
regularization. Estimator
\eqref{eq:mc_problem} was proposed in~\cite{CP10},
and solved centrally whereby all data $\cP_{\Omega_n}(\bY_n)$
is available to feed an e.g., off-the-shelf semidefinite programming (SDP) solver.
The general estimator in (P1) reduces to \eqref{eq:mc_problem}
upon setting $\bR=\mathbf{0}_{L\times F}$ and $\lambda_1=0$.
Hence, it is possible to derive a \textit{distributed} algorithm for
low-rank matrix completion by specializing Algorithm \ref{tab:table_1}
to the setting here.

Before dwelling into the algorithmic details,
a brief parenthesis is in order to touch upon properties of local
sampling operators. Operator $\cP_{\Omega_n}$ is a
linear orthogonal projector, since it projects its matrix argument onto the \emph{subspace}
$\Psi_n:=\{\bZ\in\mathbb{R}^{L_n\times T}: \supp(\bZ) \in \Omega_n\}$
of matrices with support contained in $\Omega_n$.  Linearity of $\cP_{\Omega_n}$ implies
that $\vec(\cP_{\Omega_n}(\bZ))=\bA_{\Omega_n}\vec(\bZ)$, where
$\bA_{\Omega_n}\in \mathbb{R}^{L_n\times T}$ is a symmetric
and idempotent projection
matrix that will prove handy later on. To characterize $\bA_{\Omega_n}$,
introduce an $L_n\times T$ masking matrix $\bm{\Omega}_n$ whose $(l,t)$-th
entry equals one when $(l,t) \in \Omega_n$, and zero otherwise. Since
$\cP_{\Omega_n}(\bZ)=\bm{\Omega}_n\odot \bZ$, from standard properties
of the $\vec(\cdot)$ operator it follows that
$\bA_{\Omega_n} = \textrm{diag}(\vec(\bm\Omega_n))$.

\noindent\textbf{Distributed Matrix Completion (DMC) Algorithm.}
Going back to the general distributed formulation in (P4), since there
is no sparse component $\bA$ in the matrix completion problem \eqref{eq:mc_problem},
the only constraints that remain are $\bQ_n=\bQ_m$, $m\in\cJ_n$, $n\in\cN$.
These correspond to the dual variables $\bO_n[k]$ per agent, and are
updated as shown in Algorithm~\ref{tab:table_4}.

In the absence of
$\{\bA_n\}_{n\in\cN}$ and the auxiliary variables $\{\bB_n\}_{n\in\cN}$,
it suffices to cycle over two groups of primal variables
to arrive at the DMC iterations.
The primal variable updates can be readily obtained by capitalizing
on the properties of the $\textrm{vec}(\cdot)$ operator. In
particular, Algorithm \ref{tab:table_1} indicates that the recursions
for $\bQ_n$ are given by [let $\bbq:=\textrm{vec}(\bQ')$]
\begin{align}\label{eq:Q_opt_dmc}
\nonumber\bQ_n[k+1]=\arg\min_{\bQ}&\left\{
\frac{1}{2}\|\cP_{\Omega_n}(\bY_n - \bL_n[k]\bQ')\|_{F}^{2} +
\frac{\lambda_{*}}{2N}\|\bQ\|_F^2 \right.\nonumber\\
&\left.+\langle \bO_n[k],\bQ\rangle+c\sum_{m\in\cJ_n}
\left\|\bQ-\frac{\bQ_n[k]+\bQ_m[k]}{2}\right\|_F^{2}\right\} \nonumber\\
=\textrm{unvec}\bigg(\arg\min_{\bbq}&\left\{
\frac{1}{2}\|\bA_{\Omega_n}\vec(\bY_n) - \bA_{\Omega_n} (\bI \otimes \bL_n[k]) \bbq\|^{2} +
\frac{\lambda_{*}}{2N}\|\bbq\|^2\right.\nonumber\\
&\left.+\langle \vec({\bO_n}'[k]), \bbq\rangle+c\sum_{m\in\cJ_n}
\left\|\bbq-\frac{\vec({\bQ_n}'[k] + {\bQ_m}'[k])}{2}\right\|^{2}\right\}\bigg).
\end{align}
Likewise, $\bL_n$ is updated by solving the following subproblem per iteration
(let $\bbl:=\textrm{vec}(\bL)$)
\begin{align}\label{eq:L_opt_dmc}
\bL_n[k+1]&=\arg\min_{\bL}\left\{\frac{1}{2}\|\cP_{\Omega_n}(\bY_n
- \bL\bQ_n'[k+1])\|_{F}^{2} + \frac{\lambda_{*}}{2}\|\bL\|_F^2 \right\}\nonumber\\
&=\textrm{unvec}\left(\arg\min_{\bbl}\left\{\frac{1}{2}\|\bA_{\Omega_n}\vec(\bY_n)
- \bA_{\Omega_n}(\bQ_n[k+1] \otimes \bI_{L_n} )\bbl\|^{2}
+ \frac{\lambda_{*}}{2}\|\bbl\|^2 \right\}\right).
\end{align}
Both \eqref{eq:Q_opt_dmc} and \eqref{eq:L_opt_dmc} are unconstrained
convex quadratic problems, which admit the closed-form solutions tabulated under
Algorithm~\ref{tab:table_4}.

\begin{algorithm*}[t]
\caption{: DMC algorithm per agent $n\in\cN$} \small{ \label{tab:table_4}
\begin{algorithmic}
	\STATE \textbf{Input} $\bY_n, \bA_{\Omega_n}, \lambda_{*}, c, \mu$

    \STATE \textbf{Initialize}
$\bO[0]=\mathbf{0}_{T\times \rho}$, and $\bL_n[1],\:\bQ_n[1]$ at random
    \FOR {$k=1,2$,$\ldots$}
        \STATE Receive $\{\bQ_m[k]\}$ from neighbors $m\in\cJ_n$
        \STATE {\bf [S1]} \textbf{Update local dual variables:}

        \STATE $\bO_n[k]=\bO_n[k-1] + \mu \sum_{m\in\cJ_n}(\bQ_n[k]-\bQ_m[k])$

        \STATE {\bf [S2]} \textbf{Update first group of local primal variables:}
        \STATE
        $\bE_n[k+1] = \left\{ (\bI_T \otimes \bL_n'[k])\bA_{\Omega_n} (\bI_T\otimes {\bL_n}[k]) +
        (\lambda_*/N + 2 c |\cJ_n|) \bI_{\rho T}\right\}^{-1}$
        \STATE
        $\bQ_n'[k+1]= \unvec \left( \bE_n[k+1] ~ \left\{ (\bI_T \otimes \bL_n'[k]) \bA_{\Omega_n}
        \vec(\bY_n) - \vec(\bO_n'[k]) + c  \vec(\sum_{m\in\cJ_n}(\bQ_n'[k] +
        \bQ_m'[k])) \right\} \right)$
        \STATE {\bf [S3]} \textbf{Update second group of local primal variables:}
        \STATE $\bD_n[k+1] = \left\{(\bQ_n'[k+1] \otimes \bI_{L_n}) \bA_{\Omega_n} (\bQ_n[k+1] \otimes
\bI_{L_n}) + \lambda_* \bI_{\rho L_n} \right\}^{-1}$
        \STATE $\bL_n[k+1]= \unvec\left( \bD_n[k+1] ~ (\bQ_n'[k+1] \otimes \bI_{L_n})
         \bA_{\Omega_n} \vec(\bY_n) \right)$
        \STATE Broadcast $\{\bQ_n[k+1]\}$ to neighbors $m\in\cJ_n$
    \ENDFOR
    \STATE \textbf{Return} $\bQ_n, \bL_n$
\end{algorithmic}}
\label{tab:table_4}
\end{algorithm*}

The per-agent computational complexity of the DMC algorithm is
dominated by repeated inversions of $\rho \times \rho $ and
$\rho L_n\times \rho L_n$ matrices to obtain $\bE_n[k+1]$ and
$\bD_n[k+1]$, respectively (see Algorithm \ref{tab:table_4}). Notice that
$\bE_n[k+1]\in\mathbb{R}^{\rho T \times \rho T}$ has block-diagonal
structure with blocks of size $\rho \times \rho$. Inversion of
$\rho \times \rho$ matrices is affordable in practice since $\rho$ is
typically small for a number of applications of interest
(cf. the low-rank assumption). In addition, $L_n$ is the number of
row vectors acquired per agent which can be controlled by the
designer to accommodate a prescribed maximum computational complexity. On a per iteration basis, network agents communicate
their updated local estimates $\bQ_n[k]$ only with their neighbors,
in order to carry out the updates of primal and dual variables during
the next iteration. In terms of communication cost, $\bQ_n[k]$ is a
$T \times \rho$ matrix and its transmission does not incur significant
overhead for small values of $\rho$. Observe that the dual variables $\bO_n[k]$
need not be exchanged, and the overall communication cost does not depend
on the network size $N$.


\subsection{Distributed sparse linear regression for spectrum cartography}
\label{ssec:dlasso}
In a classical linear regression setting, training data $\{\bY,\bR\}$
related through $\bY=\bR\bA + \bV$ are given and the goal is to
estimate the regression coefficient matrix $\bA$ based on the LS criterion\footnote{Even
though vector responses and model parameters are typically considered,
i.e., $\bby=\bR\bba+\bbv$,
the matrix notation is retained here for consistency.}.
However, LS  often yields unsatisfactory
prediction accuracy and fails to provide parsimonious estimates;
see e.g.,~\cite{elements_of_statistics}. To deal with such limitations, one can adopt a regularization technique known as Lasso~\cite{tibshirani_lasso}.

The general framework in this paper can accommodate distributed
estimators of the regression coefficients via Lasso, when the training
data are scattered across different agents, and their
communication to a central processing unit is prohibited for e.g.,
communication cost or privacy reasons. With reference to the
setup described in Section~\ref{sec:model}, each agent has
available the training data $\{\bY_n,\bR_n\}$ and wishes to find
\begin{align}
\hat{\bA}_{\textrm{Lasso}}:=\arg\min_{\bA} \sum_{n=1}^N\left[\frac{1}{2} \|\bY_n-\bR_n\bA\|_F^2 +
\frac{\lambda_1}{N} \|\bA\|_1\right] \label{eq:lasso}
\end{align}
in a distributed fashion. The Lasso estimator \eqref{eq:lasso} is a special case of
(P1) in the absence of a low-rank component; that is when $\bX=\mathbf{0}_{L \times T}$, and $\Omega = \{1,\ldots,L\}\times \{1,\ldots,T\}$.

An application domain where this problem arises is spectrum
sensing in cognitive radio (CR) networks whereby sensing CRs
collaborate to estimate the radio-frequency power spectrum density maps $\Phi(\bbx,f)$
across space $\bbx\in\mathbb{R}^2$ and frequency $f$;
see~\cite{juan_sensig,mateos_dlasso}. These maps enable identification of
opportunistically available spectrum bands for re-use and handoff operation;
as well as localization, transmit-power estimation, and tracking of primary user activities.

A cooperative approach to \textit{spectrum cartography} is introduced in
\cite{juan_sensig}, based on a basis
expansion model of $\Phi(\bbx,f)$. Spatially-distributed CRs collect
smoothed periodogram samples $\bY_n$ of the received signal at given sampling frequencies,
based on which they want to determine the unknown expansion coefficients in $\bA$.
The sensing scheme capitalizes on two forms of sparsity: the first one introduced by the
narrow-band nature of transmit-PSDs relative to the broad swaths
of usable spectrum; and the second one emerging from sparsely located
active radios in the operational space.
All in all, locating the active transmitters
boils down to a variable selection problem, which motivates
well employment of the Lasso in \eqref{eq:lasso}. Since data are collected by cooperating CRs at
different locations, estimation of $\bA$ amounts to
solving a distributed parameter estimation problem which demands
taking into account the network topology, and devising
a protocol to share the data. All these are accomplished by the algorithm
outlined next.

\noindent\textbf{Distributed Lasso (DLasso) Algorithm~\cite{mateos_dlasso}.}
In the Lasso setting here, (P4) reduces to
\begin{align}
\min_{\{\bA_n,\bB_n\}}&
\sum_{n=1}^N\left[\frac{1}{2}\|\bY_n
- \bR_n\bB_n\|_{F}^{2} +
\frac{\lambda_1}{N}\|\bA_n\|_1\right] \nonumber\\
\text{s. to} &\quad \bB_n=\bA_n,\quad n\in\mathcal{N}\nonumber\\
&\quad\bA_n=\bA_m, \quad
m\in\cJ_n,\:n\in\mathcal{N}\nonumber
\end{align}
which is a convex optimization problem equivalent to \eqref{eq:lasso}.
Accordingly, the DLasso algorithm (tabulated under Algorithm~\ref{tab:table_5})
is readily obtained from Algorithm \ref{tab:table_1}, after retaining only
the update recursions for $\{\bA_n,\bB_n\}$ and the multipliers $\{\bM_n,\bP_n\}$.
A distributed protocol to select $\lambda_1$ via $K$-fold cross
validation~\cite[Ch. 7]{elements_of_statistics} is also available
in~\cite{mateos_dlasso}, along with numerical tests to assess its performance.

Since Lasso in~\eqref{eq:lasso} and its separable constrained reformulation
are convex optimization problems, the general convergence results
for AD-MoM iterations can be invoked to establish convergence
of DLasso as well. A detailed proof of the following proposition can be found in
\cite[App. E]{mateos_dlasso}.

\begin{proposition}
\label{prop:convergence_DLasso} Under (a1) and for any value of the
penalty coefficient $c>0$, the iterates $\bA_n[k]$
converge to the Lasso solution [cf. \eqref{eq:lasso}] as
$k\to\infty$, i.e.,
$\lim_{k\to\infty}\bA_n[k]=\hat{\bA}_{\textrm{Lasso}},{\:}\forall{\:}n\in\cN.$
\end{proposition}

\begin{algorithm}[t]
\caption{: DLasso per agent $n\in\cN$} \small{ \label{tab:table_5}
\begin{algorithmic}
	\STATE \textbf{input} $\bY_n, \bR_n, \lambda_1, c, \mu$
    \STATE \textbf{initialize}
$\bM_n[0]=\bP_n[0]=\bA_n[1]=\bB_n[1]=\mathbf{0}_{F\times T}$
    \FOR {$k=1,2$,$\ldots$}
        \STATE Receive $\{\bA_m[k]\}$ from neighbors $m\in\cJ_n$
        \STATE {\bf [S1]} \textbf{Update local primal variables:}
                \STATE $\bM_n[k]=\bM_n[k-1]+c(\bB_n[k]-\bA_n[k])$
                \STATE $\bP_n[k]=\bP_n[k-1]+c\sum_{m\in\cJ_n}(\bA_n[k]-\bA_m[k])$
        \STATE {\bf [S2]} \textbf{Update primal variables:}
        \STATE $\bA_n[k+1]=[c(1+2|\cJ_n|)]^{-1} \cS_{\lambda_1/N}
\left(\bM_n[k]+c\bB_n[k]-\bP_n[k]+c\sum_{m\in\cJ_m}(\bA_n[k]+\bA_m[k])\right)$
		\STATE {\bf [S3]} \textbf{Update auxiliary local primal variables:}
        \STATE $\bB_n[k+1]=[\bR_n' \bR_n + c\bI_F]^{-1} \left\{\bR_n'\bY_n  -
        \bM_n[k]+ c\bA_n[k+1] \right\}$
        \STATE Broadcast $\bA_n[k+1]$ to neighbors $m\in\cJ_n$
    \ENDFOR
    \RETURN $\bA_n$
\end{algorithmic}}
\label{tab:table_5}
\end{algorithm}


\section{Numerical Tests}
\label{sec:sims}
This section corroborates convergence and gauges performance of the
proposed algorithms, when tested on the applications of Section~\ref{sec:appl} using synthetic and real
network data.

\begin{figure}[h]
\centering
  \centerline{\epsfig{figure=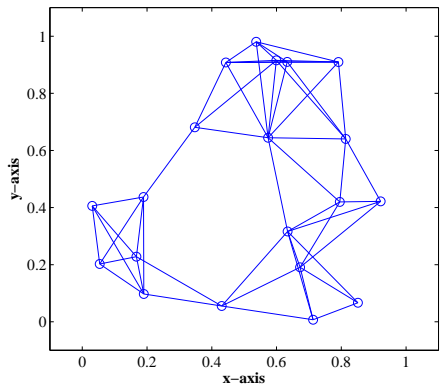,width=0.4\textwidth}}
\vspace{-5mm}\caption{A network of $N=20$ agents.}
  \label{fig:fig_1}
\end{figure}

\noindent\textbf{Synthetic network data.} A network
of $N=20$ agents is considered as a realization
of the random geometric graph model, that is, agents are randomly placed
on the unit square and two agents
communicate with each other if their Euclidean distance is less than a
prescribed communication range of $0.35$; see
Fig.~\ref{fig:fig_1}. The network graph is bidirectional and
comprises $L=106$ links, and $F=N(N-1)=380$ OD flows.
The entries of $\bV$ are independent and identically distributed (i.i.d.),
zero-mean, Gaussian with variance $\sigma^2$; i.e., $v_{l,t}\sim N(0,\sigma^2)$.
Low-rank matrices with rank $r$ are
generated from the bilinear factorization model $\bX_0 = \bW\bZ'$, where
$\bW$ and $\bZ$ are $L\times r$
and $T \times r$ matrices with i.i.d. entries drawn from Gaussian
distributions $N(0,100/F)$ and $N(0,100/T)$,
respectively. Every entry of $\bA_0$ is randomly drawn from the
set $\{-1,0,1\}$ with ${\rm Pr} (a_{i,j}=-1)={\rm Pr}(a_{i,j}=1)=\pi/2$. Unless otherwise stated, $r=3$, $\rho=3$ and $T=F=380$ are
used throughout. Different values of $\sigma$, and $\pi$ are examined.

\noindent\textbf{Internet2 network data.} Real data including
OD flow traffic levels and end-to-end
latencies are collected from the operation of the Internet2 network
(Internet backbone network across  USA)~\cite{Internet2}.
Both versions of the Internet2 network, referred as v1 and v2, are considered. OD flow
traffic levels are recorded for a three-week operation of Internet2-v1
during Dec. 8--28, 2008~\cite{LCD04},
and are used to assess performance of DUNA and DRPCA (see
Sections~\ref{subsec:sim_duna} and~\ref{subsec:sims_rpca} next).
Internet2-v1 contains $N=11$ agents, $L=41$
links, and $F=121$ flows. To test the DMC
algorithm, end-to-end flow latencies are collected from the operation
of Internet2-v2 during Aug. 18--22, 2011~\cite{Internet2}. The Internet2-v2
network comprises $N=9$ agents, $L=26$ links, and $F=81$ flows.

\noindent \textbf{Selection of tuning parameters.} The sparsity- and rank-controlling parameters
$\lambda_1$ and $\lambda_{\ast}$ are tuned to optimize performance.
The optimality conditions for (P1) indicate that for $\lambda_1 > \|\bR'\bY\|_{\infty}$
and $\lambda_{\ast} > \|\bY\|$,
$\{\bX_0=\mathbf{0}_{L \times T},\bA_0=\mathbf{0}_{F \times T}\}$
is the unique optimal solution. This in turn confines the search
space for $\lambda_1$ and $\lambda_{\ast}$ to the intervals $(0,\|\bR'\bY\|_{\infty}]$
and $(0, \|\bY\|]$, respectively. In addition, for the case of matrix completion
and robust PCA one can use the heuristic rules proposed in e.g.,~\cite{CP10} and~\cite{CLMW09}.

\subsection{Unveiling network anomalies}
\label{subsec:sim_duna}
\begin{figure}[h]
\begin{minipage}[b]{0.45\linewidth}
  \centering
  \centerline{\includegraphics[width=\linewidth, height=2 in]{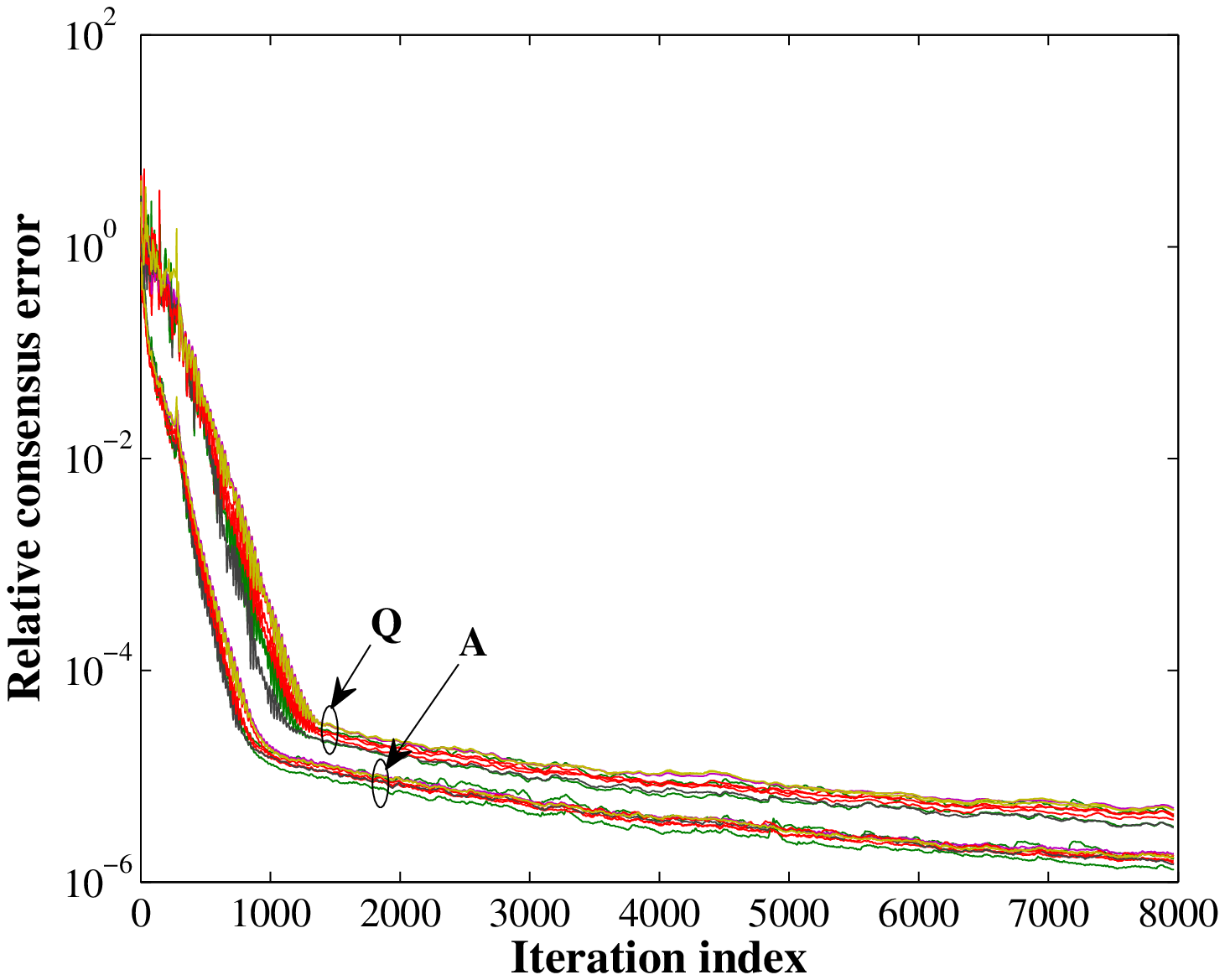}}
\medskip
\end{minipage}
\hfill
\begin{minipage}[b]{.45\linewidth}
  \centering
  \centerline{\includegraphics[width=\linewidth, height=2 in]{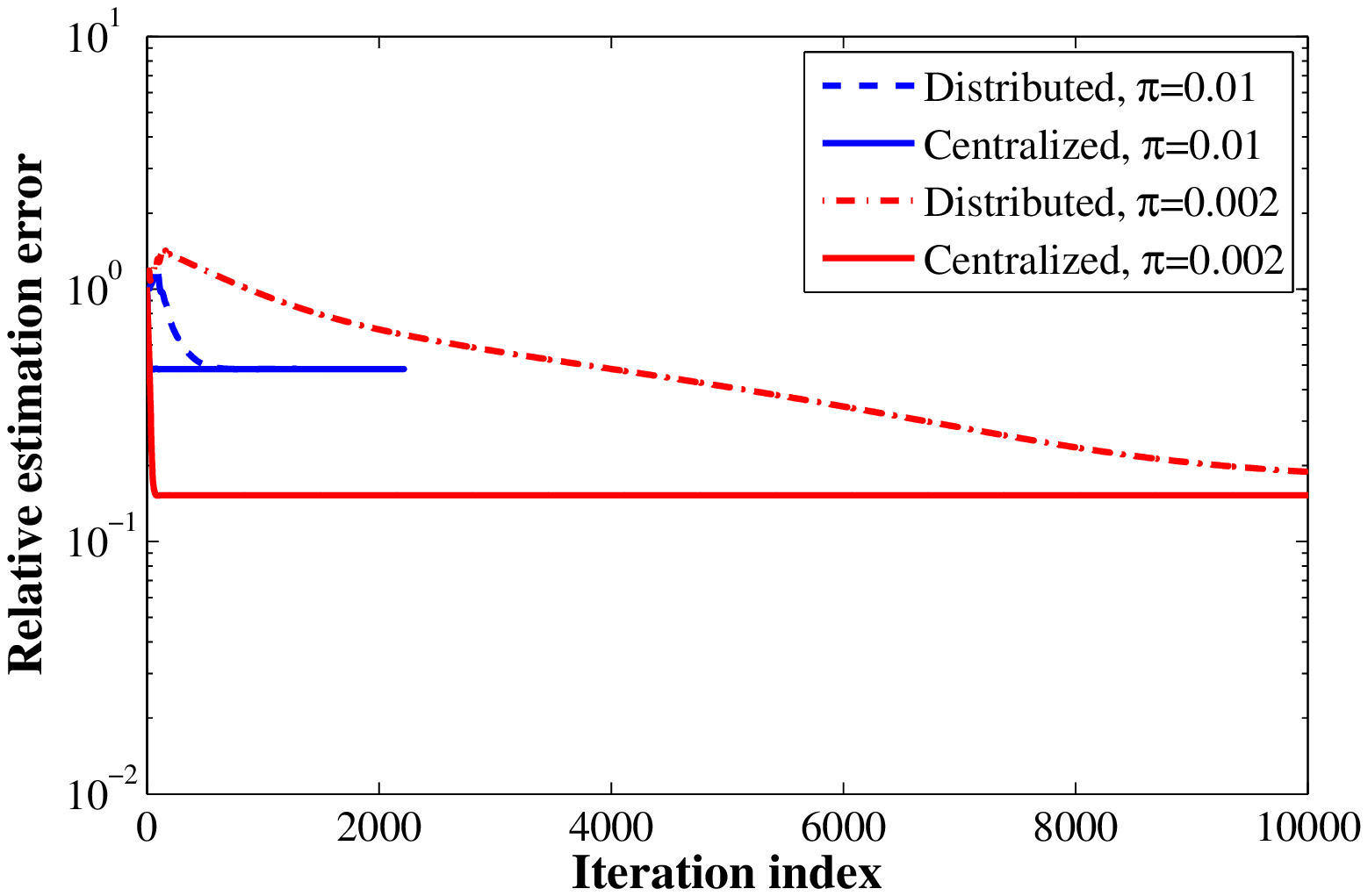}}
\medskip
\end{minipage}
\vspace{-0.7cm}
\caption{Performance of DUNA. (left) Relative consensus error for representative network
agents with $\sigma=0.01$ and $\pi=0.01$. (right) Relative estimation error for distributed and
centralized algorithms under various sparsity levels.}
\label{fig:fig_2} 
\end{figure}

Data is generated from $\bY=\bR(\bX_0+\bA_0)+\bV$, where the routing
matrix $\bR$ is obtained after determining
shortest-path routes of the OD flows. For $\mu=c=0.1$, DUNA is run until
convergence is attained. These values were
experimentally chosen to obtain the fastest convergence rate. The time evolution
of consensus among agents is depicted in Fig.~\ref{fig:fig_2} (left), for representative agents in the network. The
metric of interest here is
the relative error $\|\bQ_n[k]-\bar{\bQ}[k]\|_F/\|\bar{\bQ}[k]\|_F$
per agent $n$, which compares the corresponding local estimate with
the network-wide average $\bar{\bQ}[k]:=\frac{1}{N}\sum_{n=1}^N\bQ_{n}[k]$;
and likewise for the $\bA_n[k]$. Fig.~\ref{fig:fig_2} (left) shows that DUNA converges and
agents consent on the global matrices $\{\bQ,\bA\}$ as $k\to\infty$.

To corroborate that DUNA attains the centralized performance,
the accelerated proximal gradient algorithm of~\cite{MMG12}
is employed to solve (P1) after collecting all the per-agent
data in a central processing unit. For both the distributed and centralized schemes,
Fig.~\ref{fig:fig_2} (right) depicts the evolution of the relative error
$\|\hat{\bA}[k]-\bA_0\|_F/\|\bA_0\|_F$ for various sparsity levels, where
$\hat{\bA}[k]:=\bar{\bA}[k]$ for DUNA. It is apparent that the
distributed estimator approaches the performance of its centralized counterpart,
thus corroborating convergence and global optimality as per Proposition~\ref{th:th_1}.

\noindent\textbf{Unveiling Internet2-v1 network anomalies from SNMP measurements.}
Given the OD flow traffic measurements discussed at the
beginning of Section~\ref{sec:sims}, the link loads in $\bY$ are obtained
through multiplication with the Internet2-v1 routing matrix~\cite{Internet2}. Even though $\bY$
is ``constructed'' here from flow measurements, link loads can be typically
acquired from simple network management protocol (SNMP) traces~\cite{MC03}.
The available OD flows are a superposition of ``clean'' and anomalous traffic, i.e.,
the sum of unknown ``ground-truth'' low-rank and a sparse matrices $\bX_{0}+\bA_0$
adhering to \eqref{eq:Y} when $\bR=\bI_F$.
Therefore, the proposed algorithms are applied first to obtain a reasonably precise estimate of the
``ground-truth'' $\{\bX_0,\bA_0\}$. The estimated $\bX_0$ exhibits three dominant singular values, confirming the low-rank property of $\bX_0$.

\begin{figure}[t]
\begin{minipage}[b]{0.45\linewidth}
  \centering
  \centerline{\includegraphics[width=\linewidth, height=2 in]{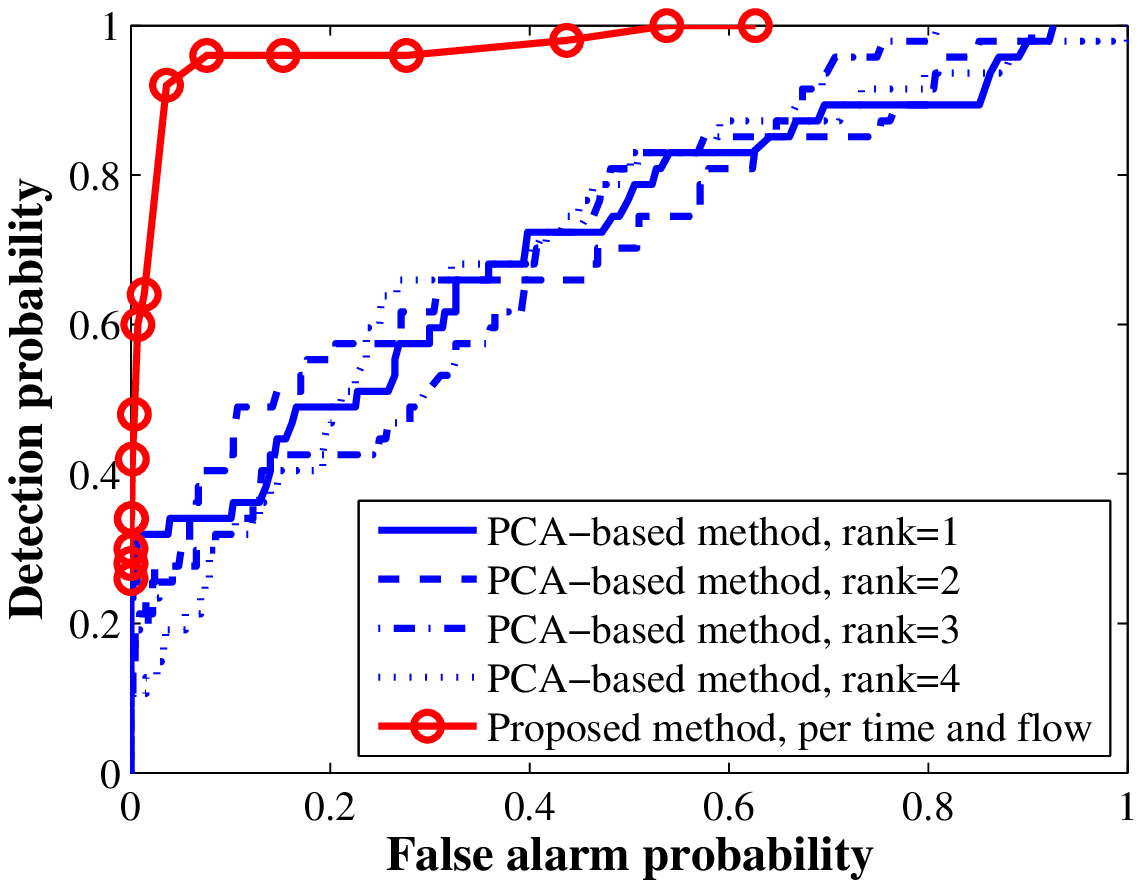}}
\medskip
\end{minipage}
\hfill
\begin{minipage}[b]{.45\linewidth}
  \centering
  \centerline{\includegraphics[width=\linewidth, height=2 in]{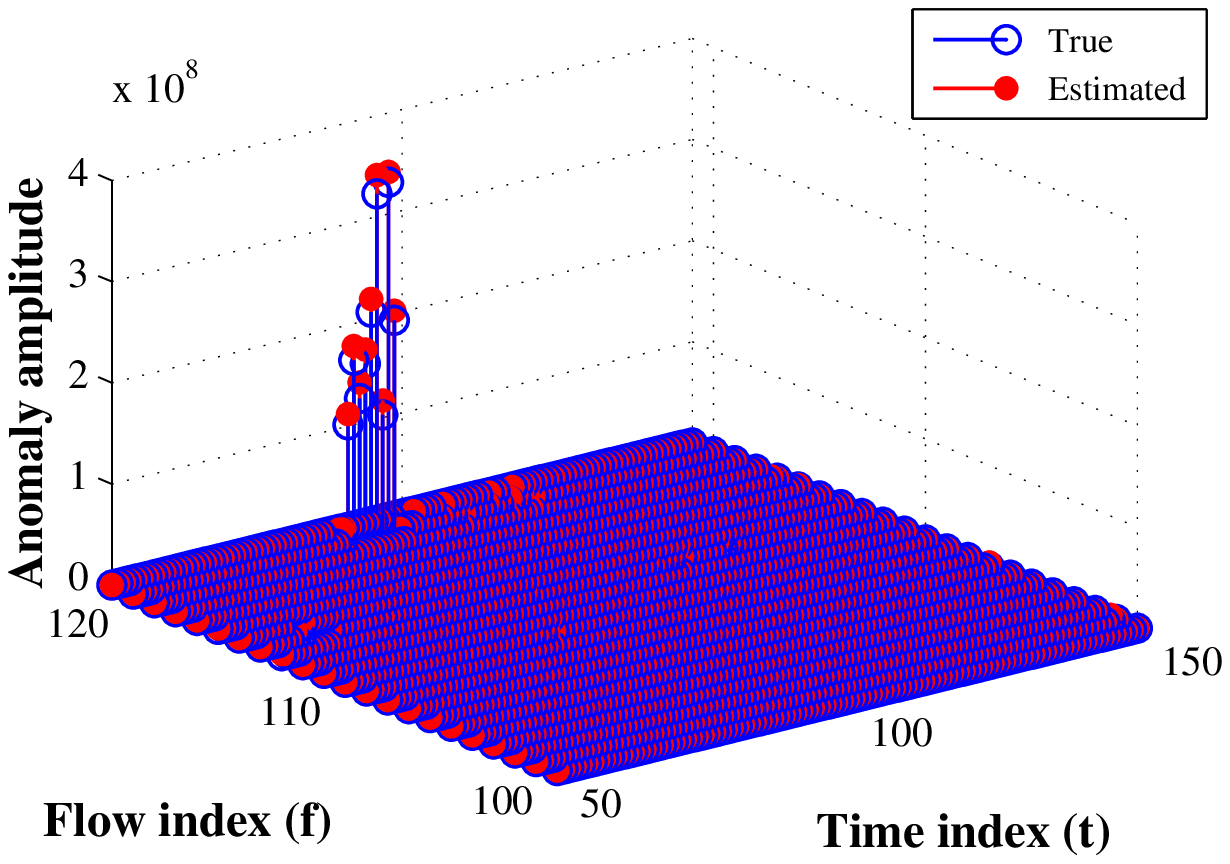}}
\medskip
\end{minipage}
\vspace{-0.7cm}
\caption{Unveiling anomalies from Internet2-v1 SNMP data. (left)
    ROC curves of the proposed versus the PCA-based method. (right)
Amplitude of the true and estimated anomalies for $\rho=5$, $P_{\rm FA}=0.04$ and $P_{\rm D}=0.93$.
}
\label{fig:fig_3} 
\end{figure}

The receiver operation characteristic (ROC) curves in Fig.~\ref{fig:fig_3} (left) highlight
the merits of DUNA when used to identify Internet2-v1 network
anomalies. Even at low false alarm rates of e.g., $P_\textrm{FA}=0.04$, the
anomalies are accurately detected ($P_\textrm{D}=0.93$). In addition, DUNA consistently
outperforms the landmark PCA-based method of~\cite{LCD04}, and can identify anomalous flows.
Fig.~\ref{fig:fig_3} (right) illustrates the magnitude of the true and
estimated anomalies across flows and time.

\subsection{Robust PCA}
\label{subsec:sims_rpca}
\begin{figure}[h]
\begin{minipage}[b]{0.45\linewidth}
  \centering
  \centerline{\includegraphics[width=\linewidth, height=2 in]{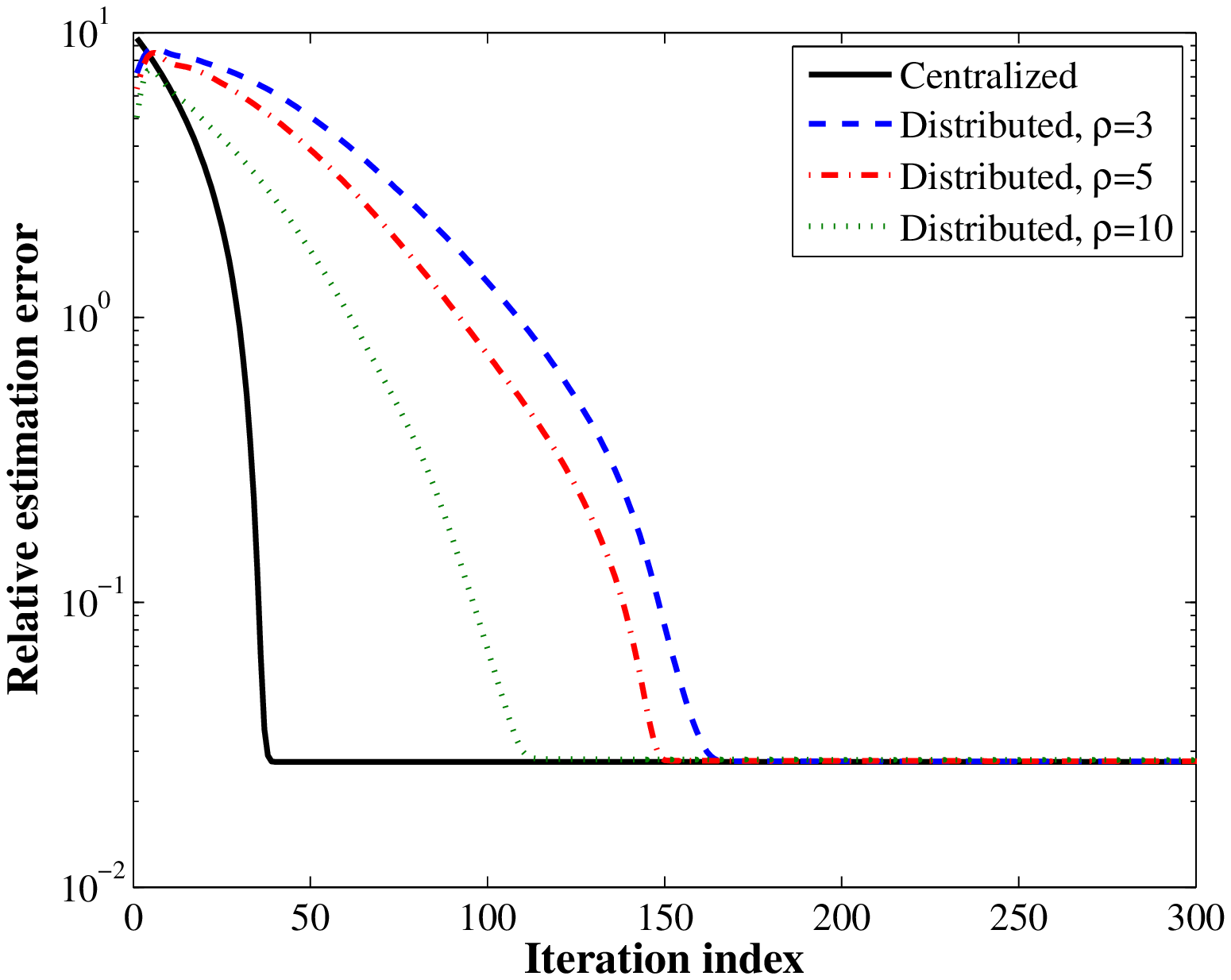}}
\medskip
\end{minipage}
\hfill
\begin{minipage}[b]{.45\linewidth}
  \centering
  \centerline{\includegraphics[width=\linewidth, height=2 in]{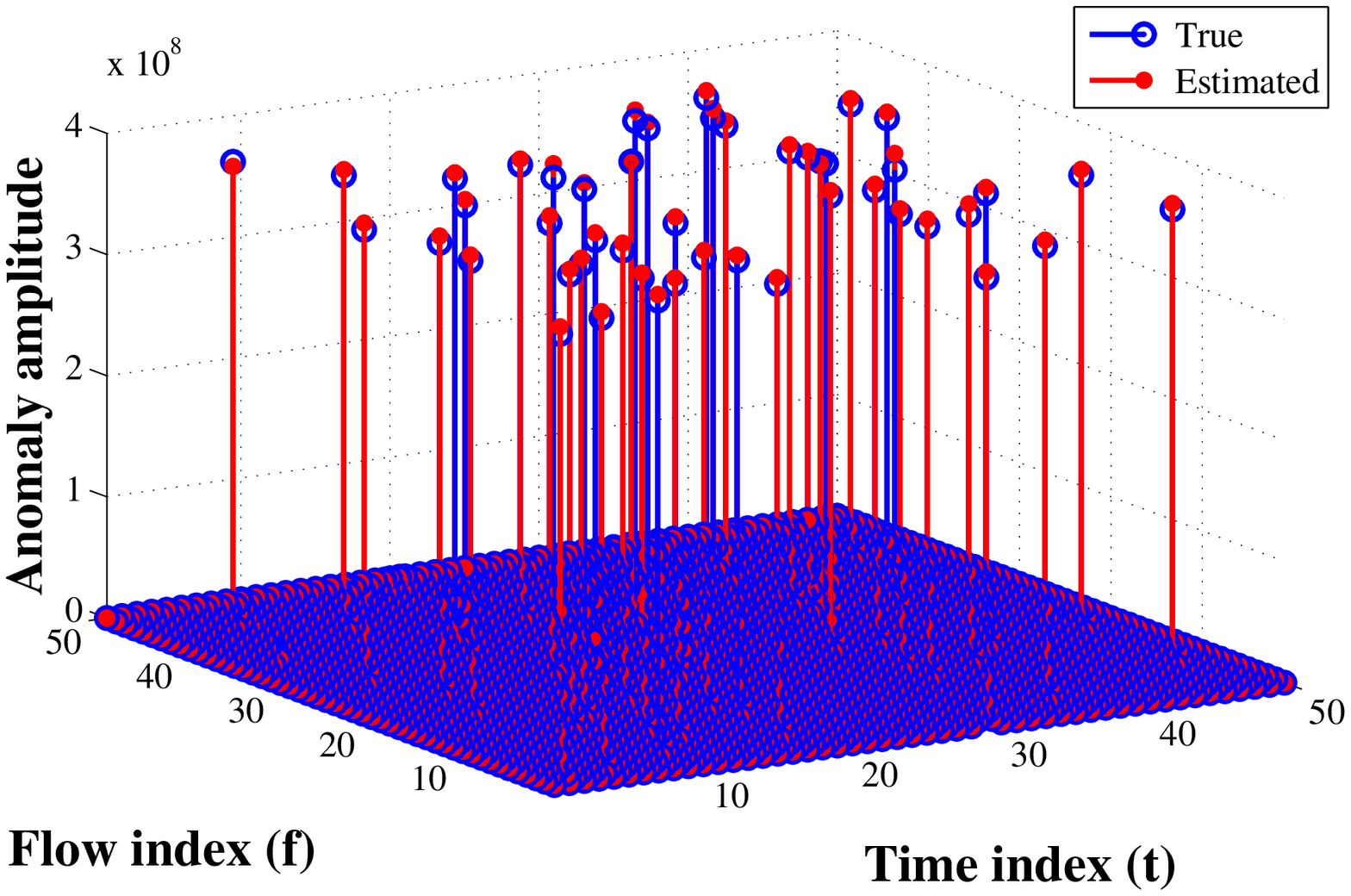}}
\medskip
\end{minipage}
\vspace{-0.7cm}
\caption{Performance of DRPCA. (left) Relative estimation error for distributed and centralized
algorithms under different $\rho$. (right) Amplitude of true and estimated anomalies using Internet2-v1 network data
when $\rho=5$, $P_{\rm FA}=10^{-3}$ and $P_{\rm D}=0.98$.
}
\label{fig:fig_4} 
\end{figure}

Next, the convergence and effectiveness of the proposed DRPCA
algorithm is corroborated with the aid of computer simulations.
An $F \times T$ data matrix is generated as $\bY=\bX_0+\bA_0+\bV$, and
the centralized estimator
\eqref{eq:rpca} is obtained using the AD-MoM method proposed in~\cite{CLMW09}. In the network setting, each agent has
available $L_n=19$ rows of $\bY$. Fig.~\ref{fig:fig_2} (right)
is replicated as Fig.~\ref{fig:fig_4} (left)
for the robust PCA problem dealt with here, and for different values of $\rho$
[the assumed upper bound on rank$(\bX_0)$].
It is again apparent that DRPCA converges and approaches the performance of \eqref{eq:rpca}
as $k\to\infty$.

\noindent\textbf{Unveiling Internet2-v1 network anomalies from Netflow measurements.}
Suppose a router $n\in\cN$ monitors the traffic volume of
OD flows to unveil anomalies using e.g., the Netflow protocol~\cite{MC03,netflowpcp}. 
Collect the time-series of all OD flows as the rows of the $F\times T$ matrix
$\bY=\bX_0+\bA_0+\bV$, where $\bA_0$ and $\bV$ account
for anomalies and noise, respectively.
As elaborated in Section~\ref{ssec:unveil_anomaly}, the common temporal patterns across flows renders the traffic matrix $\bX_0$ low-rank. Owing to the difficulties
of measuring the large number of OD flows, in practice only a few entries of
$\bY$ are typically available~\cite{MC03}, or,
link traffic measurements are utilized as in Section \ref{subsec:sim_duna} (recall that
$L\ll F$). In this example,
because the Internet2-v1 network data comprises only $F=121$ flows, it is assumed that
$\Omega=\{1,...,F\} \times \{1,...,T\}$.

To better assess performance, large spikes are injected into $1 \%$ randomly selected entries of
the ground truth-traffic matrix $\bX_0$ estimated in Section~\ref{subsec:sim_duna}.
The DRPCA algorithm is run on this Internet2-v1 Netflow data to identify
the anomalies. The results are depicted in Fig.~\ref{fig:fig_4} (right).
DRPCA accurately identifies the anomalies, achieving $P_{\rm D}=0.98$ when $P_{\rm
FA}=10^{-3}$.

\subsection{Low-rank matrix completion}
\label{sec:sims_mpc}
In addition to the synthetic data specifications outlined at the beginning
of this section, the sampling set $\Omega$ is picked uniformly
at random, where each entry of the matrix $\bm{\Omega}$ is a Bernoulli
random variable taking the value one with probability $p$. Data for
the matrix completion problem is thus generated as
$\cP_{\Omega}(\bY)=\cP_{\Omega}(\bX_0+\bV)=\bm{\Omega}\odot(\bX_0+\bV)$,
where $\bY$ is an $L \times T$ matrix with $L=T=106$. The data available to agent
$n$ is $\cP_{\Omega_n}(\bY_n)$, which corresponds to a row subset of
$\cP_{\Omega}(\bY)$.

\begin{figure}[h]
\begin{minipage}[b]{0.45\linewidth}
  \centering
  \centerline{\includegraphics[width=\linewidth, height=2 in]{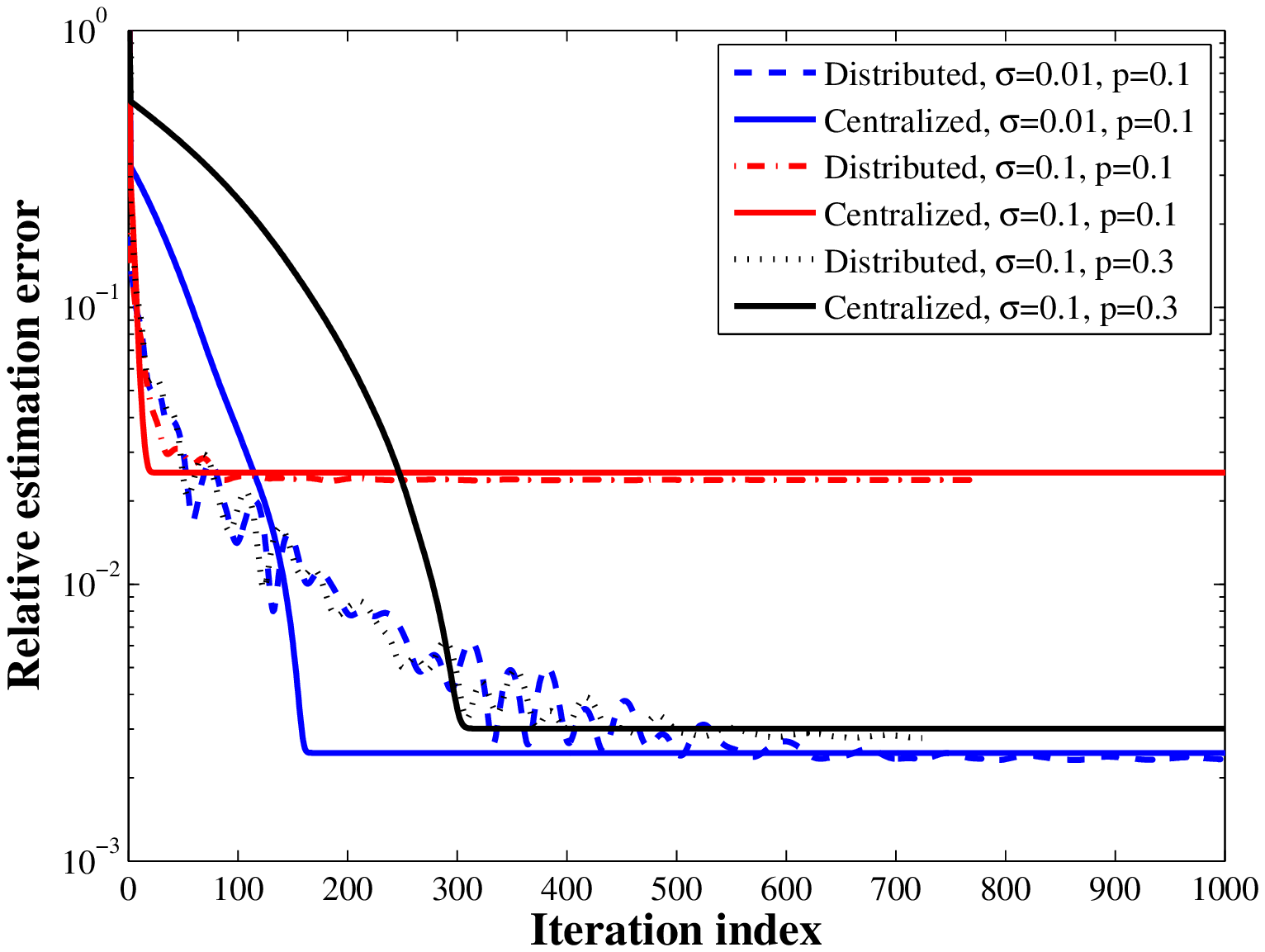}}
\medskip
\end{minipage}
\hfill
\hspace{-2mm}
\begin{minipage}[b]{.45\linewidth}
  \centering
  \centerline{\includegraphics[width=\linewidth, height=2 in]{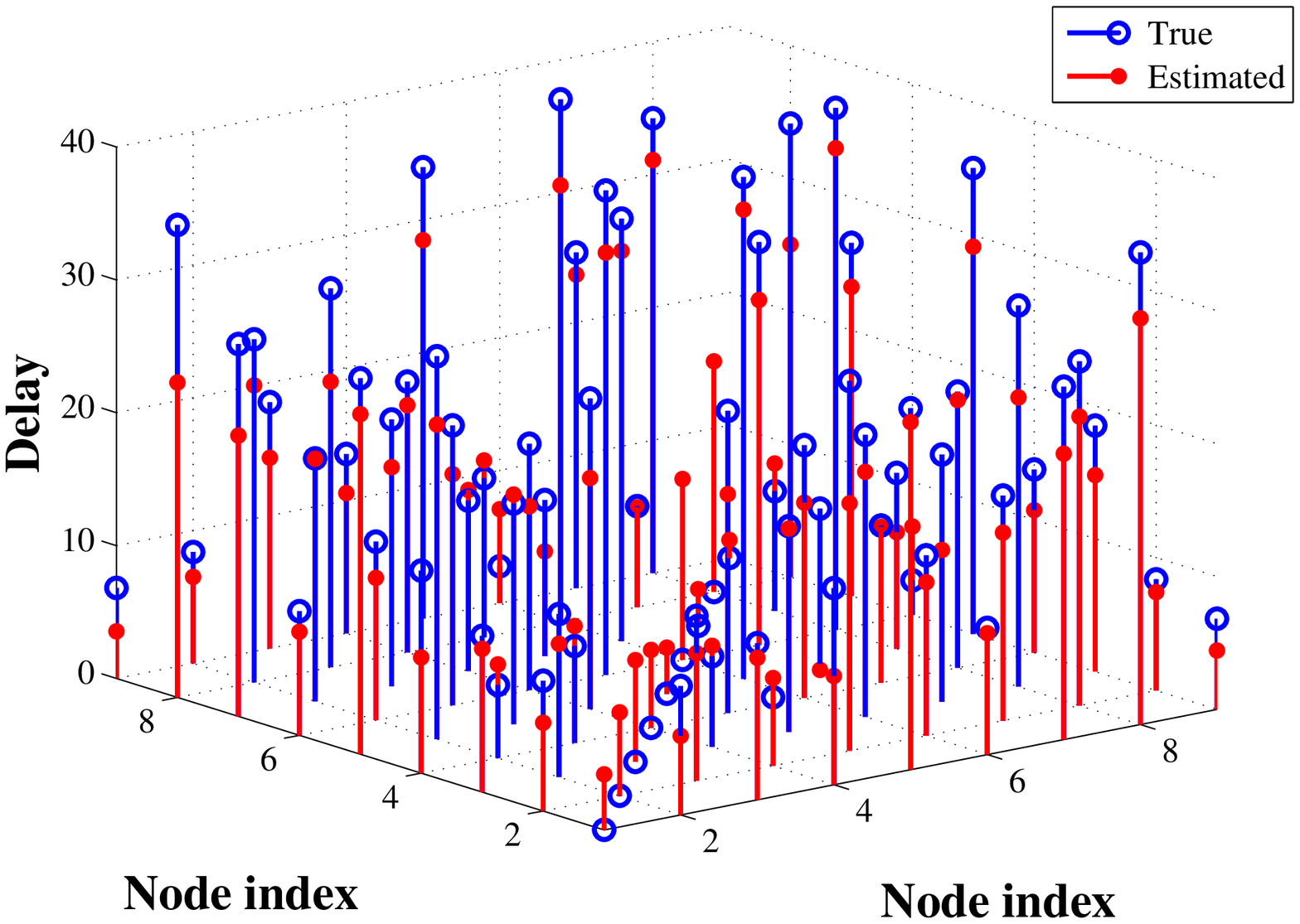}}
\medskip
\end{minipage}
\vspace{-0.7cm}
\caption{Performance of DMC. (left) Relative estimation error for distributed and centralized
algorithms under various noise strengths and percentage of missing entries. (right) Predicted and true ene-to-end
delays of Internet2-v2 network for $p=0.2$.
}
\label{fig:fig_5} 
\end{figure}

As with the previous test cases, it is shown first that the DMC algorithm
converges to the (centralized) solution
of \eqref{eq:mc_problem}. To this end, the centralized
singular value thresholding algorithm is used to solve \eqref{eq:mc_problem}~\cite{CR08},
when all data $\cP_{\Omega}(\bY)$ is available for processing.
For both the distributed and centralized schemes, Fig.~\ref{fig:fig_5} (left)
depicts the evolution of the relative error $\|\hat{\bX}[k]-\bX_0\|_F/\|\bX_0\|_F$
for different values of $\sigma$ (noise strength), and percentage of missing
entries (controlled by $p$). Regardless of the values of $\sigma$ and $p$,
the error trends clearly show the convergent
behavior of the DMC algorithm and corroborate Proposition \ref{th:th_1}.
Interestingly, for small noise levels where the estimation
error approaches zero, the distributed estimator recovers $\bX_0$ almost \emph{exactly}.

\noindent\textbf{Internet2-v2 network latency prediction.} End-to-end network
latency information is critical towards enforcing quality-of-service constraints
in many Internet applications. However, probing all pairwise delays becomes
infeasible in large-scale networks. If one collects the end-to-end latencies
of source-sink pairs $(i,j)$ in a delay matrix $\bX:=[x_{i,j}] \in \mathbb{R}^{N \times N}$,
strong dependencies among path delays render $\bX$ low-rank~\cite{latencyprediction}.
This is mainly because the paths with nearby end nodes often overlap and
share common bottleneck links. This property of $\bX$ along with the
distributed-processing requirements of large-scale networks, motivates well the
adoption of the DMC algorithm for networkwide path latency prediction.
Given the $n$-th row of $\bX$ is partially available to agent $n$, the goal
is to impute the missing delays through agent collaboration.

The DMC algorithm is tested here using the real path latency data collected from the
operation of Internet2-v2. Spectral analysis of
$\bX_0$ reveals that the first four singular values are markedly dominant,
demonstrating that $\bX_0$ is low rank. A fraction of the entries in $\bX_0$ are purposely dropped
to yield an incomplete delay matrix~$\cP_{\Omega}(\bX_0)$.
After running the DMC algorithm till convergence,
the true and predicted latencies are depicted in Fig.~\ref{fig:fig_5} (right) (for $20\%$ missing
data). The relative prediction error is around $10\%$.


\section{Concluding Summary}
\label{sec:disc}
A framework for distributed sparsity-regularized rank minimization
is developed in this paper, that is suitable for (un)supervised inference tasks in networks. Fundamental problems such as in-network
compressed sensing, matrix completion, and principal components pursuit
are all captured under the same umbrella. The novel distributed algorithms
can be utilized to unveil traffic volume anomalies from SNMP and Netflow traces, to predict networkwide path latencies in large-scale IP networks, and to map-out the RF ambiance using periodogram samples collected by spatially-distributed cognitive radios.


{\Large\appendix}


\noindent\normalsize \emph{\textbf{A. Proof of Proposition}
\ref{prop:prop_1}.} Recall the cost function of (P3) defined as
\begin{align}
f(\bL,\bQ,\bA) := \frac{1}{2}\|\cP_{\Omega}(\bY-\bL\bQ'-\bR\bA)\|_F^2 +
\frac{\lambda_*}{2}\left( \|\bL\|_F^2 +  \|\bQ\|_F^2 \right) + \lambda_1 \|\bA\|_1 \label{eq:obj_f}.
\end{align}
The stationary points $\{\bar{\bL},\bar{\bQ},\bar{\bA}\}$ of (P3) are obtained by setting to zero the (sub)gradients,
and solving~\cite{Boyd}
\begin{align}
&\partial_{\bA}{f}(\bar{\bL},\bar{\bQ},\bar{\bA})=\bR'\cP_{\Omega}(\bY-\bar{\bL}\bar{\bQ}'-\bR\bar{\bA}) -
\lambda_1 \textrm{sign}(\bar{\bA})=\mathbf{0}_{F \times T} \label{eq:df_a}\\
&\nabla_{\bL}{f}(\bar{\bL},\bar{\bQ},\bar{\bA})=\cP_{\Omega}(\bY-\bar{\bL}\bar{\bQ}'-\bR\bar{\bA})\bar{\bQ} -
\lambda_* \bar{\bL}=\mathbf{0}_{L\times \rho} \label{eq:df_l}\\
&\nabla_{\bQ'}{f}(\bar{\bL},\bar{\bQ},\bar{\bA})=\bar{\bL}'\cP_{\Omega}(\bY-\bar{\bL}\bar{\bQ}'-\bR\bar{\bA}) -
\lambda_* \bar{\bQ}'=\mathbf{0}_{\rho\times T}. \label{eq:df_q}
\end{align}
Clearly, every stationary point satisfies $\nabla_{\bL}f(\bar{\bL},\bar{\bQ},\bar{\bA})\bar{\bL}'=\mathbf{0}_{L\times L}$
and $\bar{\bQ}\nabla_{\bQ'}f(\bar{\bL},\bar{\bQ},\bar{\bA})=\mathbf{0}_{T\times T}$. Using
the aforementioned identities, the optimality conditions
\eqref{eq:df_a}-\eqref{eq:df_q}  can be rewritten as 
%
\begin{align}
&\bR'\cP_{\Omega}(\bY-\bar{\bL}\bar{\bQ}'-\bR\bar{\bA}) = \lambda_1 \textrm{sign}(\bar{\bA}) \label{eq:equality_1}\\
&\tr(\cP_{\Omega}(\bY-\bar{\bL}\bar{\bQ}'-\bR\bar{\bA})\bar{\bQ}\bar{\bL}')=
\lambda_*\tr\{\bar{\bQ}\bar{\bQ}'\}=\lambda_*\tr\{\bar{\bL}\bar{\bL}'\}.\label{eq:equality_2}
\end{align}
Consider now the following \emph{convex} optimization problem
\begin{align}
\text{(P5)}~~~\min_{\{\bX,\bA,\bW_1,\bW_2\}}& \left[\frac{1}{2}\|\cP_{\Omega}(\bY - \bX-\bR\bA)\|_{F}^{2} + \frac{\lambda_{*}}{2}\left\{\tr\{\bW_1\}+\tr\{\bW_2\}\right\} + \lambda_1
\|\bA\|_1\right]\nonumber\\
\text{s. to}\quad & \bW:=\left(\begin{array}{cc}\bW_1&\bX\\
\bX^{'}& \bW_2\end{array}\right)  \succeq \mathbf{0} \label{eq:p5_2}
\end{align}
which is \emph{equivalent} to (P1). The equivalence can be readily inferred by
minimizing (P5) with respect to $\{\bW_1,\bW_2\}$ first,
and taking advantage of the following alternative characterization of
the nuclear norm (see e.g.,~\cite{RFP07})
\begin{align}
\|\bX\|_*=\min_{\{\bW_1,\bW_2\}} ~~\frac{1}{2}\left\{\tr(\bW_1)+\tr(\bW_2)\right\},\quad
\text{s. to} \:\left(\begin{array}{cc}\bW_1&\bX\\
\bX^{'}& \bW_2\end{array}\right)  \succeq \mathbf{0}. \nonumber
\end{align}
In what follows, the optimality conditions for the conic program (P5) are explored.
To this end, the Lagrangian is first formed as
\begin{align}
\cL(\bX,\bW_1,\bW_2,\bA,\bM)=\frac{1}{2}\|\cP_{\Omega}(\bY - \bX-\bR\bA)\|_{F}^{2} +
\frac{\lambda_{*}}{2}\left\{\tr(\bW_1)+\tr(\bW_2)\right\}-\langle \bM,\bW \rangle+\|\bA\|_1 \nonumber
\end{align}
where $\bM$ denotes the dual variables associated with the conic
constraint \eqref{eq:p5_2}. For notational convenience, partition $\bM$ in four blocks $\bM_1:=[\bM]_{11}$,
$\bM_2:=[\bM]_{12}$, $\bM_3:=[\bM]_{22}$, and $\bM_4:=[\bM]_{21}$, in accordance with the block
structure of $\bW$ in \eqref{eq:p5_2}, where $\bM_1$ and $\bM_3$ are $L\times L$ and $T\times T$ matrices.
The optimal solution to (P5) must: (i) null the (sub)gradients
\begin{align}
&\nabla_{\bX}{\cL}(\bX,\bW_1,\bW_2,\bA,\bM)=-\cP_{\Omega}(\bY-\bX-\bR\bA) - \bM_2 - {\bM_4}' \label{eq:dlx}\\
&\partial_{\bA}{\cL}(\bX,\bW_1,\bW_2,\bA,\bM)=-\bR'\cP_{\Omega}(\bY-\bX-\bR\bA) -
\lambda_1\textrm{sign}(\bA) \label{eq:dla}\\
&\nabla_{\bW_1}{\cL}(\bX,\bW_1,\bW_2,\bA,\bM)=\frac{\lambda_*}{2}\bI_L - \bM_1 \label{eq:dlw_1}\\
&\nabla_{\bW_2}{\cL}(\bX,\bW_1,\bW_2,\bA,\bM)=\frac{\lambda_*}{2}\bI_T - \bM_3 \label{eq:dlw_2}
\end{align}
and satisfy (ii) the complementary slackness condition $\langle \bM,\bW \rangle=0$;
(iii) primal feasibility $\bW\succeq \mathbf{0}$; and (iv) dual feasibility $\bM \succeq \mathbf{0}$.

Recall the stationary point of (P3), and introduce
candidate primal variables $\tilde{\bX}:=\bar{\bL}\bar{\bQ}',~\tilde{\bA}:=\bar{\bA}$,
$\tilde{\bW}_1:=\bar{\bL}\bar{\bL}'$ and $\tilde{\bW}_2:=\bar{\bQ}\bar{\bQ}'$;
and the dual variables $\tilde{\bM}_1:=\frac{\lambda_*}{2}\bI_L$,
$\tilde{\bM}_3:=\frac{\lambda_*}{2}\bI_T$, $\tilde{\bM}_2:=-(1/2)\cP_{\Omega}(\bY-\bar{\bL}\bar{\bQ}'
-\bR\bar{\bA})$, and $\tilde{\bM}_4:=\tilde{\bM}_2'$.
Then, (i) holds since after plugging the candidate primal and dual variables in
\eqref{eq:dlx}-\eqref{eq:dlw_2}, the subgradients vanish. Moreover, (ii) holds since
\begin{align}
\langle \tilde{\bM},\tilde{\bW} \rangle &= \langle \tilde{\bM}_1,\tilde{\bW}_1\rangle + \langle \tilde{\bM}_2,
\tilde{\bX} \rangle + \langle \tilde{\bM}_2',\tilde{\bX}'\rangle + \langle \tilde{\bM}_3,\tilde{\bW}_2 \rangle \nonumber\\
&=\frac{\lambda_*}{2} \langle \bI_L,\bar{\bL}\bar{\bL}'\rangle + \frac{\lambda_*}{2} \langle \bL_T,\bar{\bQ}\bar{\bQ}'\rangle -
\langle \cP_{\Omega}(\bY-\bar{\bL}\bar{\bQ}'-\bR\bar{\bA}),\bar{\bL}\bar{\bQ}' \rangle \nonumber\\
&=\frac{\lambda_*}{2}\|\bar{\bL}\|_F^2 + \frac{\lambda_*}{2}\|\bar{\bQ}\|_F^2 -
 \lambda_*\|\bar{\bL}\|_F^2 =0\nonumber
\end{align}
where the last two equalities follow from \eqref{eq:equality_2}. Condition (iii) is also met since
\begin{align}
\left(\begin{array}{cc}\bar{\bL}\bar{\bL}'&\bar{\bL}\bar{\bQ}'\\
\bar{\bQ}\bar{\bL}'& \bar{\bQ}\bar{\bQ}'\end{array}\right)=\left(\begin{array}{c}\bar{\bL}\\
\bar{\bQ}\end{array}\right) \left(\begin{array}{c}\bar{\bL}\\
\bar{\bQ}\end{array}\right)' \succeq \mathbf{0}.
\end{align}
To satisfy (iv), based on a Schur complement
argument~\cite{Horn} it suffices to enforce $\sigma_{\max}(\tilde{\bM}_2) \leq \lambda_*$.
\hfill$\blacksquare$


\noindent\normalsize \emph{\textbf{B. Derivation of Algorithm \ref{tab:table_1}}.}
It is shown here that [S1]-[S4] in Section \ref{ssec:ad_mom}
give rise to the set of recursions tabulated under Algorithm \ref{tab:table_1}.
To this end, recall the augmented Lagrangian function in
\eqref{augLagr} and focus first on [S4]. From the decomposable
structure of $\mathcal{L}_c$, \eqref{S4_ADMOM} decouples into simpler strictly convex
sub-problems for $n\in\cN$ and $m\in\cJ_n$, namely
\begin{align}
\label{eq:opt_B} \bB_n[k+1]{}={}&\arg\min_{\bB_n}\left\{r_n(
\bL_n[k+1],\bQ_n[k+1],\bB_n)+\langle \bM_n[k],\bB_n\rangle
+\frac{c}{2}\|\bB_n-\bA_n[k+1]\|_F^2\right\}\\
\label{eq:opt_F}\bar{\bF}_n^m[k+1]{}={}&\mbox{arg}\:\min_{\bar{\bF}_n^m}
\left\{\frac{c}{2}\left(\|\bQ_n[k+1]-\bar{\bF}_n^m\|_F^2
+\|\bQ_m[k+1]-\bar{\bF}_n^m\|_F^2\right)
-\langle \bar{\bC}_n^m[k]+\tilde{\bC}_n^m[k],\bar{\bF}_n^m\rangle\right\}\\
\label{eq:opt_G}\bar{\bG}_n^m[k+1]{}={}&\mbox{arg}\:\min_{\bar{\bG}_n^m}
\left\{\frac{c}{2}\left(\|\bA_n[k+1]-\bar{\bG}_n^m\|_F^2
+\|\bA_m[k+1]-\bar{\bG}_n^m\|_F^2\right)
-\langle \bar{\bD}_n^m[k]+\tilde{\bD}_n^m[k],\bar{\bG}_n^m\rangle\right\}.
\end{align}
Note that in formulating \eqref{eq:opt_F} and \eqref{eq:opt_G}, the auxiliary variables $\tilde{\bF}_n^m$ and $\tilde{\bG}_n^m$ were eliminated using the constraints $\bar{\bF}_n^m=\tilde{\bF}_n^m$ and $\bar{\bG}_n^m=\tilde{\bG}_n^m$, respectively. The unconstrained quadratic problems \eqref{eq:opt_F} and \eqref{eq:opt_G} admit the closed-form solutions
\begin{align}
\bar{\bF}_n^m[k+1]{}={}&\tilde{\bF}_n^m[k+1]=\frac{1}{2c}
(\bar{\bC}_n^m[k]+\tilde{\bC}_n^m[k])
+\frac{1}{2}\left(\bQ_n[k+1]+\bQ_m[k+1]\right)\label{Flocalrecursion}\\
\bar{\bG}_n^m[k+1]{}={}&\tilde{\bG}_n^m[k+1]=\frac{1}{2c}
(\bar{\bD}_n^m[k]+\tilde{\bD}_n^m[k])
+\frac{1}{2}\left(\bA_n[k+1]+\bA_m[k+1]\right).\label{Glocalrecursion}
\end{align}
Using \eqref{Flocalrecursion} to eliminate $\bar{\bF}_n^m[k]$ and
$\tilde{\bF}_n^m[k]$ from \eqref{eq:multi_barC} and \eqref{eq:multi_tildeC} respectively, a simple induction argument establishes that if the initial Lagrange multipliers obey $\bar{\bC}_n^m[0]=-\tilde{\bC}_n^m[0]=\mathbf{0}_{T\times\rho}$, then $\bar{\bC}_n^m[k]=-\tilde{\bC}_n^m[k]$ for all $k\geq 0$, where $n\in\cN$ and $m\in\cJ_n$. Likewise, the same holds true for $\bar{\bD}_n^m[k]$ and $\tilde{\bD}_n^m[k]$. The collection of multipliers $\{\tilde{\bC}_n^m[k],\tilde{\bD}_n^m[k]\}_{n\in\cN}^{m\in\cJ_n}$ is thus redundant, and \eqref{Flocalrecursion}-\eqref{Glocalrecursion} simplify to
\begin{align}
\bar{\bF}_n^m[k+1]{}={}&\tilde{\bF}_n^m[k+1]=
\frac{1}{2}\left(\bQ_n[k+1]+\bQ_m[k+1]\right),\:\:n\in\cN,\:\: m\in\cJ_n\label{Flocalrecursion_simple}\\
\bar{\bG}_n^m[k+1]{}={}&\tilde{\bG}_n^m[k+1]=
\frac{1}{2}\left(\bA_n[k+1]+\bA_m[k+1]\right),\:\:n\in\cN,\:\: m\in\cJ_n.\label{Glocalrecursion_simple}
\end{align}
Observe that $\bar{\bF}_n^m[k]=\bar{\bF}_m^n[k]$ and $\bar{\bG}_n^m[k]=\bar{\bG}_m^n[k]$  for all $k\geq0$, identities that will be used later on. By plugging \eqref{Flocalrecursion_simple} and \eqref{Glocalrecursion_simple} into \eqref{eq:multi_barC} and \eqref{eq:multi_barD} respectively, the non-redundant multiplier updates become
\begin{align}
\bar{\bC}_n^m[k]&=\bar{\bC}_n^m[k-1]+\frac{\mu}{2}\left(\bQ_n[k]-\bQ_m[k]\right),
    {\quad}n\in\cN,\:m\in\cJ_n\label{eq:multi_barC_update}\\
\bar{\bD}_n^m[k]&=\bar{\bD}_n^m[k-1]+\frac{\mu}{2}\left(\bA_n[k]-\bA_m[k]\right),
    {\quad}n\in\cN,\:m\in\cJ_n\label{eq:multi_barD_update}.
\end{align}
If $\bar{\bC}_n^m[0]=-\bar{\bC}^n_m[0]=\mathbf{0}_{T\times\rho}$,
then the structure of \eqref{eq:multi_barC_update} reveals that
$\bar{\bC}_n^m[k]=-\bar{\bC}^n_m[k]$ for all $k\geq
0$, where $n\in\cN$ and $m\in\cJ_n$. Clearly, the same holds
true for $\bar{\bD}_n^m[k]$, and these identities will become handy
in the sequel.

Moving on to [S3], \eqref{S3_ADMOM} decouples into $|\cN|$ unconstrained
quadratic sub-problems
\begin{equation*}
\bL_n[k+1]=\arg\min_{\bL_n}\left\{r_n(\bL_n,\bQ_n[k+1],\bB_n[k]) +
\frac{\lambda_{*}}{2}\|\bL_n\|_F^2 \right\}.
\end{equation*}
%
The minimization \eqref{S2_ADMOM} in [S2] also decomposes into simpler
sub-problems, both across agents and across the variables $\{\bQ_n\}_{n\in\calN}$ and $\{\bA_n\}_{n\in\calN}$,
which are decoupled in the augmented Lagrangian when all other variables
are fixed. Specifically, the per agent updates of $\bQ_n$ are given by
\begin{align}\label{eq:Q_opt}
\nonumber\bQ_n[k+1]=\arg\min_{\bQ_n}&\left\{
r_n(\bL_n[k],\bQ_n,\bB_n[k]) +\frac{\lambda_{*}}{2N}\|\bQ_n\|_F^2
+\sum_{m\in\cJ_n}\langle\bar{\bC}_n^m[k]+\tilde{\bC}_m^n[k]
,\bQ_n\rangle\right.\nonumber\\
&\left.+\frac{c}{2}\sum_{m\in\cJ_n}
\left(\|\bQ_n-\bar{\bF}_n^m[k]\|_F^{2}+
\|\bQ_n-\tilde{\bF}_m^n[k]\|_F^2\right)\right\}\nonumber\\
\nonumber=\arg\min_{\bQ_n}&\left\{
r_n(\bL_n[k],\bQ_n,\bB_n[k]) +
\frac{\lambda_{*}}{2N}\|\bQ_n\|_F^2
+\langle \bO_n[k],\bQ_n\rangle\right.\nonumber\\
&\left.+c\sum_{m\in\cJ_n}
\left\|\bQ_n-\frac{\bQ_n[k]+\bQ_m[k]}{2}\right\|_F^{2}\right\}
\end{align}
where the second equality was obtained after using: i) $\bar{\bC}_n^m[k]=\tilde{\bC}_m^n[k]$ which follows from the identities
$\bar{\bC}_n^m[k]=-\tilde{\bC}_n^m[k]$ and $\bar{\bC}_n^m[k]=-\bar{\bC}_m^n[k]$ established earlier; ii) the definition $\bO_n(k):=2\sum_{m\in\cJ_n}\bar{\bC}_n^m[k]$; and iii) the identity
$\bar{\bF}_n^m[k]=\tilde{\bF}_m^n[k]$, which allows to merge the identical quadratic penalty terms and eliminate both $\bar{\bF}_n^m[k]$ and
$\tilde{\bF}_m^n[k]$ using \eqref{Flocalrecursion_simple}. 

Upon defining $\bP_n(k):=2\sum_{m\in\cJ_n}\bar{\bD}_n^m[k]$ and following
similar steps as the ones that led to the second equality in \eqref{eq:Q_opt},
one arrives at
\begin{align}\label{eq:A_opt}
\nonumber\bA_n[k+1]=\arg\min_{\bA_n}&\left\{
\frac{\lambda_1}{N}\|\bA_n\|_1-\langle \bM_n[k],\bA_n\rangle
+\frac{c}{2}\|\bB_n[k]-\bA_n\|_F^2+\langle \bP_n[k],\bA_n\rangle\right.\nonumber\\
&\left.+c\sum_{m\in\cJ_n}
\left\|\bA_n-\frac{\bA_n[k]+\bA_m[k]}{2}\right\|_F^{2}\right\}\nonumber\\
=\arg\min_{\bA_n}&\left\{\frac{N}{2}\left\|\bA_n
-\frac{\bM_n[k]+c\bB_n[k]-\bP_n[k]+c\sum_{m\in\cJ_n}\left(\bA_n[k]+\bA_m[k]
\right)}
{c(1+2|\cJ_n|)}\right\|_F^{2}+\lambda_1\|\bA_n\|_1\right\}
\end{align}
where the second equality follows by completing the squares. Problem
\eqref{eq:A_opt} is a separable instance of the Lasso (also related
to the proximal operator of the $\ell_1$-norm); hence, its
solution is expressible in terms of the soft-thresholding operator
as in Algorithm \ref{tab:table_1}.


\noindent\normalsize \emph{\textbf{C. Proof of Proposition~\ref{th:th_1}}.} Let
$\bar{\bQ}_n:=\lim_{k\to\infty}\bQ_n[k]$, and likewise for all other convergent
sequences in Algorithm \ref{tab:table_1}.
Examination of the recursion for $\bO_n[k]$ in the limit
as $k\to\infty$, reveals that $\sum_{m\in\cJ_n}[\bar{\bQ}_n-\bar{\bQ}_m]=\mathbf{0}_{T\times\rho}, \
\forall\: n\in\cN$. Upon vectorizing the matrix quantities involved,
this system of equations implies that the
supervector $\bar{\mathbf{q}}:=[\textrm{vec}[\bar{\mathbf{Q}}_1]',\ldots,
\textrm{vec}[\bar{\mathbf{Q}}_N]']'$ belongs to the
nullspace of $\bbL\otimes\mathbf{I}_{T\rho}$, where $\mathbf{L}$ is
the Laplacian of the network graph
$G(\cN,\cL)$. Under (a1), this guarantees
that $\bar{\mathbf{Q}}_1=\bar{\mathbf{Q}}_2=\ldots=\bar{\mathbf{Q}}_N$.
From the analysis of the limiting behavior of $\bP_n[k]$,
the same argument leads to
$\bar{\mathbf{A}}_1=\bar{\mathbf{A}}_2=\ldots=\bar{\mathbf{A}}_N$, which establishes
the consensus results in the statement of Proposition \ref{th:th_1}.
Hence, one can
go ahead and define $\bar{\mathbf{Q}}:=\bar{\mathbf{Q}}_n$ and
$\bar{\mathbf{A}}:=\bar{\mathbf{A}}_n$. Before moving on, note that
convergence of $\bM_n[k]$ implies that
$\bar{\mathbf{B}}_n=\bar{\mathbf{A}}_n=\bar{\mathbf{A}}$, $n\in\cN$.
These observations guarantee that the limiting solution is feasible for (P4).

To prove the optimality claim it suffices to show that upon convergence,
the fixed point $\{\bar{\bL},\bar{\bQ},\bar{\bA},\bar{\bB}\}$ of the iterations comprising
Algorithm \ref{tab:table_1} satisfies the Karush-Kuhn-Tucker (KKT) optimality
conditions for (P4). Proposition~\ref{prop:prop_1} asserts
 that if $\|\cP_{\Omega}(\bY-\bL\bQ'-\bR\bA)\|\leq\lambda_{\ast}$,
$\{\bar{\bL},\bar{\bQ},\bar{\bA}\}$ is indeed an optimal solution to (P1).
To this end, consider the updates of the primal variables in Algorithm~\ref{tab:table_1},
which satisfy
\begin{align}
&\nabla_{\bQ_n}r_n(\bQ_n[k+1],\bL_n[k],\bB_n[k]) + \frac{\lambda_{\ast}}{N} \bQ_n[k+1] + \bO_n[k+1] \nonumber\\
&\hspace{45mm}+ 2c \sum_{m\in\cJ_n}\left(\bQ_n[k+1]-\frac{\bQ_n[k]+\bQ_m[k]}{2}\right) =
\mathbf{0}_{T\times\rho}\label{eq:grad_qn}\\
&\nabla_{\bL_n}r_n(\bQ_n[k+1],\bL_n[k+1],\bB_n[k]) + \lambda_{\ast} \bL_n[k+1] =
\mathbf{0}_{L\times\rho}\label{eq:grad_ln}\\
&\nabla_{\bB_n}r_n(\bQ_n[k+1],\bL_n[k+1],\bB_n[k+1]) + \bM_n[k] + c(\bB_n[k+1]-\bA_n[k+1]) = \mathbf{0}_{F\times
T}.\label{eq:grad_bn}
\end{align}
Taking the limit from both sides of \eqref{eq:grad_qn}--\eqref{eq:grad_bn},
and summing up over all $n\in\cN$ yields
\begin{align}
&\nabla_{\bQ}r(\bar{\bQ},\bar{\bL},\bar{\bA}) + \lambda_{\ast} \bar{\bQ} = \mathbf{0}_{T\times\rho}
\label{eq:limit_fq}\\
&\nabla_{\bL}r(\bar{\bQ},\bar{\bL},\bar{\bA}) + \lambda_{\ast} \bar{\bL} = \mathbf{0}_{L\times\rho}
\label{eq:limit_fl}\\
&\nabla_{\bB}r(\bar{\bQ},\bar{\bL},\bar{\bA}) + \sum_{n\in\cN}\bar{\bM}_n = \mathbf{0}_{F\times T} \label{eq:limit_fa}
\end{align}
where $r(\bL,\bQ,\bB):=\frac{1}{2}\|\cP_{\Omega}(\bY - \bL\bQ' - \bR\bB)\|_F^2$.
To arrive at \eqref{eq:limit_fq}, the assumption that
$\bar{\bC}_n^m[1]=\mathbf{0},~\forall m\in\cJ_n,n\in\cN$ is used,
and thus $\bar{\bC}_n^m[k]=-\bar{\bC}_m^n[k]$ which leads to $\sum_{n\in\cN}\bO_n[k]=
\sum_{n\in\cN}\sum_{m\in\cJ_n}\bar{\bC}_n^m[k]=\mathbf{0}$.

Next, consider the auxiliary matrices $\bm \Theta_n:=\bar{\bM}_n-\bar{\bP}_n+
c(1+2|\cJ_n|)\bar{\bA}$, $n\in\cN$. In the limit as $k\to\infty$, the update
recursion for $\bA_n[k+1]$ in Algorithm \ref{tab:table_1}
can be written as
$c(1+2|\cJ_n|)\bar{\bA}=\calS\left(\bm \Theta_n,\lambda_1/N\right).$
Proceed by defining $\bm\Psi_n:=\bm
\Theta_n-c(1+2|\cJ_n|)\bar{\bA}$, and observe that
the input-output relationship of the soft-thresholding operator
$\calS$ yields
\begin{equation}\label{delta_properties}
[\bm\Psi_n]_{f,t}=\left\{%
\begin{array}{ll}
    \lambda_1/N, & [\bar{\bA}]_{f,t}>0, \\
    -\lambda_1/N, & [\bar{\bA}]_{f,t}<0, \\
    \xi^{(n)}_{f,t} :|\xi^{(n)}_{f,t}|\leq\lambda_1/N , & [\bar{\bA}]_{f,t}=0.
  \end{array}%
\right.
\end{equation}
Given \eqref{delta_properties}, define the
matrices $\bm \Gamma_1:=\frac{1}{2}\left(
\lambda_1\mathbf{1}_F\mathbf{1}_T'+\sum_{n=1}^N \bm \Psi_n\right)$ and
$\bm \Gamma_2:=\frac{1}{2}\left(
\lambda_1\mathbf{1}_F\mathbf{1}_T'-\sum_{n=1}^N \bm \Psi_n\right)$,
and show that
they satisfy the following properties: (i) $\bm \Gamma_1, \bm \Gamma_2\geq \mathbf{0}$
(entrywise); (ii) $[\bm \Gamma_1]_{f,t}=0,$
if $[\bar{\bA}]_{f,t}<0$; (iii) $[\bm \Gamma_2]_{f,t}=0,$ if $[\bar{\bA}]_{f,t}>0$;
(iv) $\bm \Gamma_1+\bm \Gamma_2=\lambda_1\mathbf{1}_F\mathbf{1}_T'$; and (v)
$\bm \Gamma_1-\bm \Gamma_2=\sum_{n\in\cN}\bar{\bM}_n$.
Properties (i)-(iii) follow after
adding up the result in \eqref{delta_properties} for $n=1,2,\ldots,N$.
Property (iv)
is readily checked from the definitions of $\bm \Gamma_1$ and
$\bm \Gamma_2$. Finally, (v) follows since
\begin{equation*}
\bm \Gamma_1-\bm \Gamma_2=\sum_{n=1}^N \bm \Psi_n=\sum_{n=1}^N
\left(\bm
\Theta_n-c(1+2|\cJ_n|)\bar{\bA}\right)=
\sum_{n=1}^N\bar{\bM}_n-\sum_{n=1}^N\bar{\bP}_n=\sum_{n=1}^N\bar{\bM}_n
\end{equation*}
where $\sum_{n=1}^N\bar{\bP}_n=\mathbf{0}$ (from the
identity $\sum_{n=1}^N\bP_n[k]=\mathbf{0}$) is used to obtain the last equality.

The proof is concluded by noticing that properties (i)-(v) along with
\eqref{eq:limit_fq}-\eqref{eq:limit_fa} comprise the
KKT conditions for the following optimization problem
\begin{align}
\min_{\{\bL,\bQ,\bA,\bT\}}~& \frac{1}{2}\|\cP_{\Omega}(\bY - \bL\bQ' - \bR\bA)\|_{F}^{2} +
\frac{\lambda_{*}}{2}\left\{\|\bL\|_F^2 + \|\bQ\|_F^2 \right\}+\lambda_1\mathbf{1}_{F}'
\bT\mathbf{1}_{T}  \nonumber\\
\text{s. to}~&-\bT\leq \bA \leq \bT \quad \textrm{(entrywise)}\nonumber
\end{align}
where $\{\bar{\bL},\bar{\bQ},\bar{\bA}\}$ and $\{\bm \Gamma_1,\bm \Gamma_2\}$
 play the role of the optimal primal
and dual variables, respectively. This last problem is clearly equivalent to
(P4).\hfill$\blacksquare$


\bibliographystyle{IEEEtranS}
\bibliography{IEEEabrv,biblio}


\vspace{-1cm}

\vspace{-2cm}

\vspace{-3cm}

\vspace{-3cm}

\end{document}